\documentclass[journal,10pt]{IEEEtran}

\usepackage{epsfig}
\usepackage{multirow}
\usepackage{algorithm}
\usepackage{algorithmic}
\usepackage{clrscode}
\usepackage{microtype} %better margin management
\usepackage{booktabs}  %better looking tables
\usepackage{url}
\usepackage[caption=false]{subfig}
\usepackage{amsmath}
\usepackage{pifont}
\newcommand{\xmark}{\text{\ding{55}}}

\newcommand{\etal}{\mbox{\textit{et al.}}}

\usepackage{color}

\begin{document}

\title{Physical Design Obfuscation of Hardware: A Comprehensive Investigation of Device- and Logic-Level Techniques}

\author{Arunkumar~Vijayakumar\IEEEauthorrefmark{1},~\IEEEmembership{Student~Member,~IEEE,} Vinay~C.~Patil\IEEEauthorrefmark{1},~\IEEEmembership{Student~Member,~IEEE,} Daniel~E.~Holcomb\IEEEauthorrefmark{1},~\IEEEmembership{Member,~IEEE,} Christof~Paar\IEEEauthorrefmark{1}\IEEEauthorrefmark{2},~\IEEEmembership{Fellow,~IEEE,} and~Sandip~Kundu\IEEEauthorrefmark{1},~\IEEEmembership{Fellow,~IEEE}

\thanks{Manuscript received on ...}
\thanks{\IEEEauthorrefmark{1}The authors are with the Deparment of Electrical and Computer Engineering, University of Massachusetts Amherst 01002, USA (email: \{avijayak, vcpatil, dholcomb, cpaar, kundu\}@umass.edu)}
\thanks{\IEEEauthorrefmark{2}Prof. Christof Paar is also affiliated with Horst G{\"o}rtz Institute for IT-Security, Ruhr-Universit{\"a}t Bochum, Germany (email: christof.paar@rub.de)}
\thanks{This work is supported in part by grants from NSF (grants no. 1421352 and 1563829), ERC (grant no. 695022), and Intel.}
}

\maketitle

\begin{abstract}
The threat of hardware reverse engineering is a growing concern for a large number of applications. 
A main defense strategy against reverse engineering is hardware obfuscation. 
In this work we investigate physical obfuscation techniques, which perform alterations of circuit elements that are difficult or impossible for an adversary to observe. Examples of such stealthy manipulations are changes in the doping concentrations or dielectric manipulations. An attacker will, thus,  extract a netlist which does not correspond to the logic function of the device-under-attack. This approach of camouflaging has garnered recent attention in the literature. 

In this paper, we expound on this promising direction to conduct a systematic end-to-end study of the VLSI design process to find multiple ways to obfuscate a circuit for hardware security. 
This paper makes three major contributions. First, we provide a categorization of the available physical obfuscation techniques as it pertains to various design stages. There is a large and multidimensional design space for introducing obfuscated elements and mechanisms, and the proposed taxonomy is helpful for a systematic treatment. Second, we provide a review of the methods that have been proposed or in use. Third, we present recent and new device and logic-level techniques for design obfuscation. For each technique considered, we discuss feasibility of the approach and assess  likelihood of its detection. Then we turn our focus to open research questions, and conclude with suggestions for future research directions.

\end{abstract}

\section{Introduction}

The threat of hardware reverse engineering is a growing concern for a large number of applications, from consumer electronics and cyber-physical systems all the way to military systems. The are two broad goals an adversary might pursue: (1) learning the functionality of the device-under-attack and (2) gaining knowledge that enables manipulation of the target. With respect to the first goal, reverse engineering of ICs may be motivated by a number of reasons. For example, the adversary may wish to steal IP for their own use, or to resell IP without licensing. They may also wish to analyze a device to gain a competitive advantage or to leapfrog technology. 
Regarding the second goal, active design manipulations, a particularly attractive target for attackers are implementations of hardware-based security functions such as cryptographic algorithms or random number generators. 
An attacker may want to learn the details of a security module with the goal of subsequently weakening it or inserting carefully crafted Trojans. In addition to manipulating security modules, an adversary can also be interested in adding or disabling functionality to commercial ICs \cite{Rambus_Personalization}. Note, however, that a reverse engineer does not necessarily need to pursue illegal or ethically questionable objectives. An ``adversary'' in the reverse engineering setting can also represent a stakeholder attempting to discover IP violations or faulty product designs. Furthermore, reverse engineering for competitive reasons is legal in most Western countries, including the United States and the European Union~\cite{torrance-11}. A further discussion of adversary goals is given in Section~\ref{sec:attacker_model}.

The problem of hardware reverse engineering is not only theoretical or limited to powerful adversaries. Numerous hardware attacks over the last few years have shown the ease of removing coating and de-layering circuits~\cite{nohl-08,DBLP:conf/crypto/StrobelDKLOSP13,kammerstetter-14-breaking,skorobogatov2012physical,Skorobogatov:2012:BSS:2413316.2413319}. Moreover reverse engineering is also increasingly supported by CAD tools such as Chipworks' ICWorks and the open-source tool Degate~\cite{schobert-thesis}. In addition, reverse engineering equipment such as scanning electronic microscopes and Focused Ion Beam (FIB) have become considerably more accessible over the past decade due to wide-spread availability in research labs and a drop in price for new and used equipment.

An IC designer who wants to prevent reverse engineering has a difficult task because the adversary will have physical possession of the chip. Since physical anti-tamper measures to prevent invasive attacks are technically and economically infeasible in most applications, the designer's main line of defense is obfuscation of the hardware. There are two principal approaches to hardware obfuscation: \textit{structural obfuscation} and what we coin \textit{physical design obfuscation}\footnote{We would like to clarify the terminologies: The terms physical design/physical mechanism used in this paper refer to device and interconnect-level techniques that can be used for obfuscation. It does not refer to physical design methods used for placement and routing of circuits.}. 

Structural obfuscation --- which tries to hide the true functionality of the device-under-attack  through techniques such as randomized placement of logic elements, irregular routing or dummy wires --- is widely used in industry. Unfortunately, structural obfuscation does not prevent the adversary from recovering the complete netlist through de-layering. With structural obfuscation, she faces the task of learning the functionality from the recovered netlist which has a structure that makes this very task more complicated. However, given enough resources and time, an attacker will most likely succeed. This situation is analogous to software reverse engineering. In both cases, the designer does not have a true advantage over the attacker, i.e., the situation is symmetric in the sense that the adversary can observe whatever methods the designer employs. 

This paper addresses the second family of techniques, physical design obfuscation, which performs alterations of circuit elements which are \textit{difficult or impossible to observe} for the adversary. Examples of such stealthy manipulations are localized changes in doping concentrations, dielectric manipulation, and open connections in the sub-nanometer range. In contrast to ``classical'' structural obfuscation,  when stealthy manipulations are used, the adversary will extract a netlist which does \textit{not} unambiguously correspond to the logic function of the device under analysis (DUA). Gate camouflaging is one such approach to physical design obfuscation \cite{rajendran-13,cocchi2014cktcamo}. Gate camouflaging and similar techniques are promising, as they give the designer a true advantage of knowledge over the adversary. For a successful reverse engineering attack, the adversary has to recover both the basic design structure and also the stealthy manipulations; the former is easy to recover while the latter should be hard. This situation is akin to classical cryptography in which an cryptographic algorithm is known while the key is secret and must be learned by the attacker. Due to the asymmetry of the setting which gives the designer an advantage, this approach to obfuscation has the potential to provide strong protection.

Despite the promise of this approach, there is scant treatment in the literature and industry on this topic. The contribution at hand attempts to give a comprehensive treatment to the emerging and promising area of physical design obfuscation. The main contributions of this paper are as follows:
\begin{itemize}
	\item We present a taxonomy of physical design obfuscation techniques (Sec.~\ref{sec:taxonomy}).
	\item We investigate existing and novel physical design obfuscation techniques at the device level (Sec.~\ref{sec:physical_mechanisms})
	\item We investigate novel methods for employing physical design obfuscation at the logic-level  (Sec.~\ref{sec:logical_mechanisms}) by utilizing the device-level techniques.
\end{itemize}
To the best of our knowledge, this is the first time that the topic of physical design manipulations for obfuscation is taxonomically surveyed. We believe this work will be valuable for industry, where ad-hoc methods dominate the issue of circuit protection through obfuscation. The proposed work also has the potential to initiate further research related to physical design mechanisms, as well as research in realizing complex circuits  and evaluating them with respect to resistance to reverse engineering.

The paper is organized as follows.  Section~\ref{sec:attacker_model} discusses the adversary setting and Section~\ref{sec:taxonomy} introduces a taxonomy for physical design obfuscation. In Section~\ref{ssec:prev_work} we summarize related work.  As main contributions, we discuss in Sections~\ref{sec:physical_mechanisms} and \ref{sec:logical_mechanisms} the various physical manipulations and logic-level mechanisms, respectively, that can be used for obfuscation. 
The paper concludes with a discussion of open challenges in Section~\ref{sec:discussion}.

\section{Adversary Setting} 
\label{sec:attacker_model}
Hardware obfuscation is a tactic used by a designer to hide functionality from an unauthorized third party. In order to systematically treat the somewhat murky question of hiding, two aspects about adversaries need to be addressed. First, what are the possible objectives of an attacker, and second, what are her capabilities? The answers to these questions influence how successful a given obfuscation technique can be. 

We note that the chip manufacturer (foundry) has perfect information about the layout on all levels, including the details about physical obfuscation methods. Hence, a manufacturer can, at least in theory, understand circuit functionality despite all physical obfuscation methods that may be employed. However, in most civilian applications, the designer's primary concern is not an adversary who is in cohort with a foundry. The far more common case is an adversary who is in possession of one or more obfuscated ICs, and  who attempts to extract information about the circuit. As we will see below, attackers can vary widely with respect to goals (which information he wants to extract) and his capabilities (what he can observe and manipulate). This class of attack is sometimes referred to as Man-at-the-End (MATE) attack in software security community \cite{Akhunzada2015mate}.

\subsection{Objectives of the Adversary}
The most obvious goal of an attacker --- to extract the complete function of the circuit realized by the device-under-attack from its physical structure ---- is only one of many. For instance, in competitive analysis, the adversary might merely be interested in learning which algorithms are being used in a given product. Such objectives are often less complex and, thus, easier to achieve.  We refer interested readers to the work of Rostami \etal~\cite{rostami2014primer} for more background on this issue. As noted earlier, a hardware reverse engineer does not necessarily need to pursue illegal or ethically questionable objectives, but she can also attempt to  discover IP violations or faulty product designs. Below we list some common goals of an adversary, roughly ordered from easiest to hardest.

\begin{enumerate}
\item Learning the device properties and design rules of fabricated ICs (process analysis). For instance, the estimated design rules for an Intel 22nm tri-gate process can be obtained by commercially-offered reverse engineering for \$16,000 USD~\cite{intel-22nm-dfm}.
\item Learning the location of sensitive signal wires or buses. This information can enable subsequent use of targeted micro-probing to extract internal data values. These values may be immediately relevant as in the case of cryptographic keys, or the values may provide more information about the design, e.g., state bits of an Finite-state machine (FSM) or register values.
\item Learning which specific IP blocks or algorithms are being used in the target design. This does not require learning the details of the implementation. Possible motivations include competitive intelligence or detection of IP violations. 
\item Learning the Boolean function of combinational logic, or the FSM of sequential logic (circuit extraction). This can apply to the entire IC or only certain parts which are proprietary and/or sensitive. The adversary's goal might be to clone or even alter the product.
\item Learning the logic function of each gate and the interconnections between them. This implies learning the overall Boolean function, but also includes additional information about exactly how the function is implemented, and the extra information could enable side channel attacks etc. The incentives for this are the same as in the previous point, and an additional incentive is for performing detailed competitive analysis. 
\item Learning the GDSII of an entire IC or of parts, e.g., an IP block. This will enable the adversary to re-manufacture an identical IC in the same technology. This is the highest value information that can be obtained by reverse engineering.
\end{enumerate}

\subsection{Capabilities of the Attacker}
The effectiveness of obfuscation will depend on the assumed capabilities of the attacker. Publicly available tools are available to facilitate these methods of attack~\cite{schobert-thesis}. Torrance and James~\cite{torrance-11} give an overview of the state of the art for reverse engineering in 2011. The following provides some attacker capabilities ordered roughly from weakest to strongest. 
\begin{enumerate}
\item The attacker is able to observe all I/O activities and can create and
manipulate input values at will.
\item  The attacker is able to decapsulate the IC and to probe values on block-level I/O boundaries or top-level metal layers between blocks with approximately 1k--10k gates.
\item The attacker is able to observe the state of sequential elements at any time as would occur with scan chain access; Rajendran \etal~\cite{rajendran-13} assume their attacker to have this ability.
\item  The attacker is able to delayer chip and learn all metal layers and corresponding vias but is not able to distinguish real from dummy vias; this implies ability to identify gate structures as done by Nohl \etal~\cite{nohl-08} and others.
\item The attacker is able to delayer chip and resolve sub-gate-level width and spacing of metal or polysilicon features of layers. Torrance and James~\cite{torrance-11} note that optical imaging suffices to extract such finer details at technologies above 0.18$\mu$m, and that Scanning Electron Microscope (SEM) can accomplish the same below 0.18$\mu$m if the dielectric is removed.
\item The attacker is able to detect dopant programming of transistors using techniques like passive voltage contrast~\cite{sugawara-14}.
\item The attacker is able to delayer chip and further can distinguish between vias and fake vias; Rajendran \etal~\cite{rajendran-13} and others assume an attacker that lacks this capability.
\end{enumerate}
A summary of various measurement techniques that can be used in reverse engineering is presented in Table I.

\begin{table}[th]
  \setlength{\tabcolsep}{4pt}
  \centering
  \caption{Various measurement techniques used in reverse engineering}
  \begin{tabular}{ l | l | l }
Technique                                & Capability                            &  Cost \\ \hline
PICA imaging~\cite{tsang2000pica}        & Observe transistors switching         &  High \\ \hline
SQUID Microscopy~\cite{nikawa2001squid}  & Observe constant current from shorts  &  High \\ \hline
\parbox[c][5ex]{0.4\columnwidth}{Scanning Electron Microscopy (SEM)}            & Imaging of features below 200nm~\cite{torrance-11}  & Moderate   \\ \hline
Optical Microscopy (OP)                      & Imaging of features down to 200nm~\cite{torrance-11} & Low    \\ \hline
Passive Voltage Contrast  (PVC)               & Detect Doping                         &   Low  \\ \hline
  \end{tabular}
  
  \label{tab:measurement_techniques}
\end{table}

\section{A Taxonomy for Physical Design Obfuscation}
\label{sec:taxonomy}
%\todo{maybe section needs better title}

In the literature, low-level hardware obfuscation has rarely been treated in a systematic way. At the same time, there is a large and multi-dimensional design space for introducing hardware obfuscation. In fact, there is a wealth of physical design obfuscation techniques which result in stealthy circuits, i.e., circuits whose realized logic behavior differs from the apparent logic function extracted during reverse engineering. We propose a taxonomy in order to systematically treat this multifaceted topic. In fact, many obfuscation techniques often blur the boundaries between these layers. Nevertheless, these layers are helpful for systematic organization of apparently disparate physical manipulation techniques. In normal VLSI design process, successive transformations from one abstraction layer to the next result in the final physical design. However in obfuscation, the goal is to add/subtract some information specific to cross-layer transformations. Therefore, describing obfuscation purely in terms of extracting the higher levels of abstraction from lower level is complicated and may not always be possible. For example, we can introduce layout artifacts to deliberately induce obfuscation through signal integrity which has no gate-level functional equivalent at the logic abstraction layer. It is for this reason, we introduce a different nomenclature for expressing obfuscation instead of using standard VLSI design abstractions. 

\begin{figure}[!t]
\centering
\includegraphics[width=0.9\columnwidth]{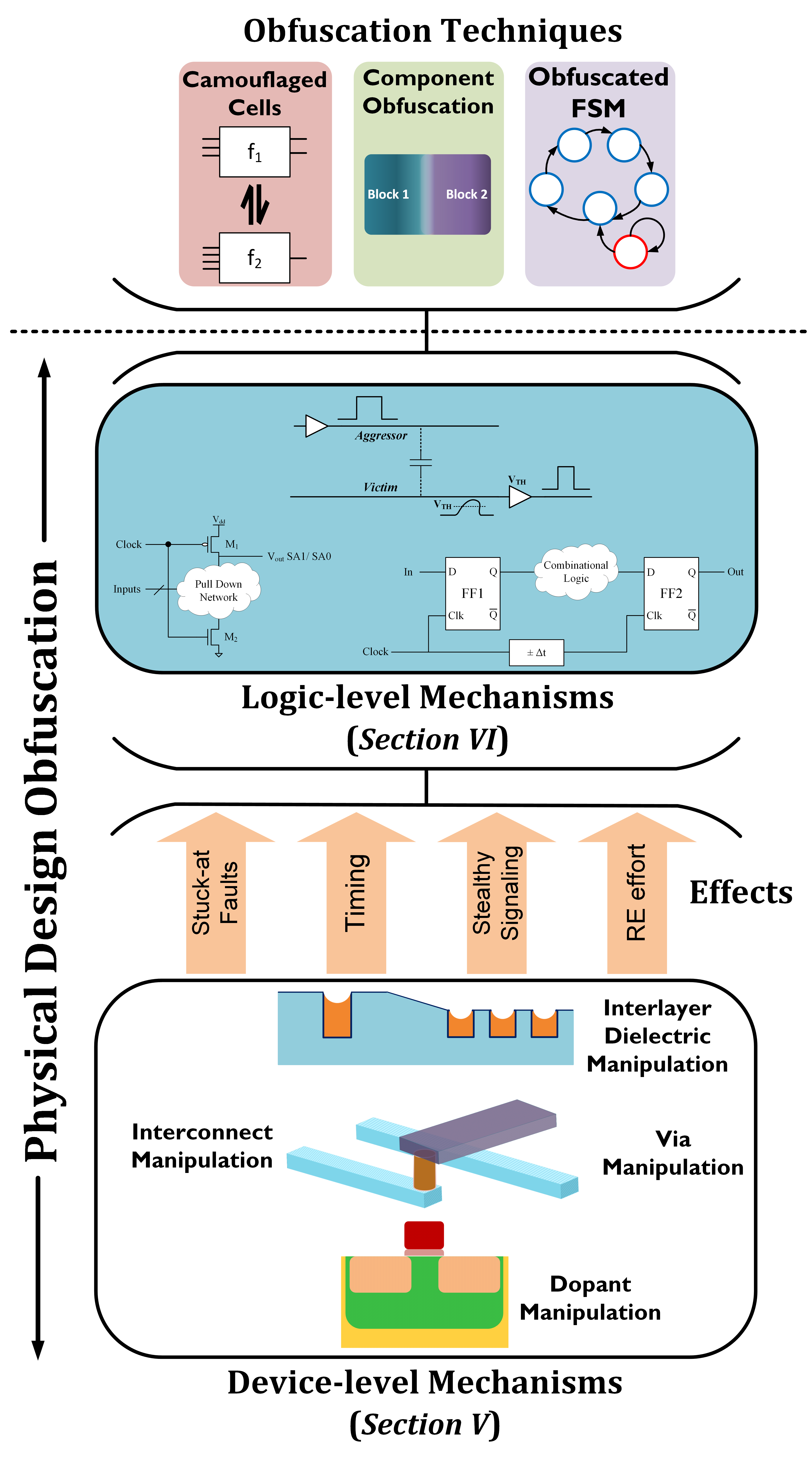}
\caption{A structured view on physical design obfuscation: device \& interconnect mechanisms on the bottom layer and logic mechanisms on the middle layer enable actual obfuscation techniques on the top layer.}
\label{fig:hier_rep}
\end{figure}

Hardware obfuscation based on stealthy circuit manipulations can be viewed in a three-layer model, shown in \figurename~\ref{fig:hier_rep}. The lower two layers are formed by (i) device and interconnect mechanisms (bottom layer) and logic-level mechanisms (middle) layer. They are in  hierarchical structure, i.e., logic-level mechanisms are based on the device and interconnect manipulations. These two layers are  forms of physical design obfuscation and are the subject of this contribution. Physical design obfuscation provide the basic mechanisms with which actual obfuscation techniques can be realized. Such  techniques form the top layer in \figurename~\ref{fig:hier_rep}. We elaborate on each of the three layers below. 

\textbf{Bottom Layer: Device-level mechanisms} refer to measures such as re-sizing of transistors or interconnect manipulations. Even though there are numerous such possible alterations, they can be classified into three groups. Each of the groups has a  certain type of \textit{effect} on the circuit, as shown in  \figurename~\ref{fig:hier_rep}: (\textit{i}) stuck-at-faults, (\textit{ii}) stealthy signaling and (\textit{iii}) delay manipulation.

\textit{Stuck-at-faults} refer to changes that cause certain circuit nodes to have a constant logical 0 or 1, even though the apparent circuit structure might imply some other function. 

\textit{Stealthy signaling} refers to situations where an adversary may be unable to tell whether physical structures are communicating, and this includes both direct connection and non-contact influencing mechanisms such as cross-talk. 

\textit{Delay manipulation} refers to techniques that make a node switch slower or faster than it would appear to, and uses this to change the sequential behavior of a chip.

\textbf{Middle Layer: Logic-level mechanisms} The logic level is concerned with circuit structures which are built from the stealthy manipulations from the bottom layer described above. In subsequent higher system layers, the logic-level elements are used as building blocks for actual obfuscation techniques to protect integrated circuits. These logic-level mechanisms form the second level of abstraction and can to some extent be independent of the specifics of the device-level mechanisms (i.e., sizing, dopant, etc). For example, one can design a circuit that uses physically obfuscated look-up tables while abstracting away the lower level, i.e.,  what physical mechanisms are actually used to create the obfuscated look-up tables.

\textbf{Top Layer: Obfuscation Techniques} The last layer of the obfuscation hierarchy is used by techniques such as gate camouflaging, component obfuscation and FSM obfuscation. These higher-layer techniques can be used to protect complex circuits and systems that are realized on an IC. There have been several interesting proposals for securing hardware using obfuscation techniques.

\quad %just an empty line

Most of the higher-level obfuscation techniques on the top layer of \figurename~\ref{fig:hier_rep} have been published relatively recently, cf. Section~\ref{ssec:prev_work}. However, there is a void regarding which underlying physical design mechanisms exist for realizing higher-level obfuscation. The focus of the contribution at hand is to provide a comprehensive treatment of the lower-level mechanisms. First, we introduce in a structured way a variety of device and interconnect mechanisms which can be used for physical design obfuscation in Section~\ref{sec:physical_mechanisms}. The second focus is on logic-level mechanisms which are discussed in Section~\ref{sec:logical_mechanisms}, i.e., techniques which belong to the middle layer in the figure.		

\section{Related Work}
\label{ssec:prev_work}
Even though hardware obfuscation has been used for a long time in commercial and military ICs, the topic had received scant treatment in the scientific community. However, over the last few years, the general area of hardware protection has become more active.  Roughly speaking, the proposed methods can be classified into system- and circuit-level techniques, and we summarize related work below. 

\subsection{System-level Techniques}
Some of the system-level methods that have been proposed can be viewed as ``classical'' obfuscation approaches, whereas others are more general techniques for circuit protection, i.e., their focus is not directly on hiding of elements. 
 
\textbf{Component Obfuscation}: Many reverse engineering techniques engage in \textit{component recognition}. Techniques have been proposed to increase the complexity of delineating components by changing the logic gates at the output and input boundaries of connected blocks (Boundary Blurring) \cite{parham2010hiding}, replacing an entire circuit with an obfuscated equivalent (Component Fusion) \cite{mcdonald2011DCV, mcdonald2009protecting} and targeting interconnections between two blocks for encryption~\cite{mcdonald2006program}.

\textbf{Obfuscation with Programmable Logic}: Wendt \etal ~propose replacing subset of gates of a circuit with Physically Unclonable Function (PUF) and FPGA-based logic \cite{wendt2014obfpuf}. The PUF is assumed to be unique and unclonable, and the connected FPGA is used to reproduce functionality of the replaced logic. This approach provides a custom solution for each chip but it comes at considerable area and run-time costs.

\textbf{FSM Modification}: Alkabani \etal \cite{alkabani2007active,alkabani2007remote} propose FSM modification techniques for the purpose of locking a chip by obfuscating its power-up state. Only the correct key input sequence allows the circuit to be initialized correctly. Li and Zhou \cite{li2013obfseq} propose a methodology for obfuscating sequential circuits and embedding a secret key in the power-up state of the IC that must be present to unlock full unthrottled performance. Chakraborty and Bhunia \cite{chakraborty2009harpoon} propose a methodology to perform simultaneous obfuscation and authentication of an SoC design netlist. 

\textbf{Related Techniques --- Logic Encryption:} Logic encryption is a relatively recent hardware security concept introduced by Roy, Koushanfar and Markov \cite{roy2008epic}. In logic encryption, additional gates are added to the design such that only the correct input values, considered the key, allow the circuit to work as intended. Wrong values can lead to corrupted output~\cite{rajendran2012logic} or even can lock the circuit. Logic encryption can be an effective tool against over production, however, it is not effective against invasive attacks. It has been shown that it is possible to reverse engineer logic encryption keys in time that is linear in the number of keys~\cite{rajendran2012sec} and quite efficient in practice using Boolean satisfiability problem (SAT) solving~\cite{subramanyan-15}. An attacker may also be able to observe the added logic and trace the key inputs by using imaging techniques.

\subsection{Circuit-level Obfuscation}
The primary goals of circuit-level obfuscation methods have been to create or modify cell libraries to hide gate functions and adding non-essential structures to the design. The underlying assumption for these approaches is that an adversary  is presented with extra complexity when reverse-engineering the logic.

\textbf{Camouflaged Cells}: Obfuscation via camouflaging of cells can be achieved through custom design to either mimic other cells or to allow for post-manufacturing programmability to customize gate functions. Relevant literature dealing with cell camouflaging can be found in patents by Baukus \etal~\cite{baukus-98-patent, baukus1999digital} and Cocchi \etal~\cite{cocchi2012building}.

\textbf{Filler Cells}: Chow \etal\  \cite{chow2012camouflaging} and Cocchi \etal\ \cite{cocchi2014cktcamo} deal with utilizing filler cells with routing in a realistic fashion to create a dense network. Some of the cells may even be connected to the functional logic but do not hamper its operation. The filler cells and their associated routing increase the amount of data that an attacker will need to sort through during reverse engineering in order to extract the underlying logic.

\subsection{Hardware Security Surveys}
Erstwhile research has resulted in extensive surveys on protecting hardware intellectual property (IP). The work by Tehranipoor and Wang \cite{Tehranipoor2011IHSbook} provides a good understanding of various aspects of hardware security and trust with focus on all types of electronic devices and systems. Colombier and Bossuet \cite{Colombier2014SurveyIPprotect} provide a detailed discussion on protection of design data and IP with exhaustive survey of previous research into hardware protection. Guin \etal \cite{Guin2014counterfeit} provide a thorough analysis of the problem of counterfeiting and examine various countermeasures including hardware obfuscation.
These papers provide a summary  of various hardware obfuscation techniques proposed in literature. In contrast, the focus of this paper is on surveying \textit{physical design manipulation mechanisms} and presenting device- and logic-level techniques for hardware obfuscation.

\section{Device-Level Mechanisms}
\label{sec:physical_mechanisms}
In this section, we discuss physical design obfuscation through device-level techniques, which we consider to be manipulations of transistors and the interconnections between them. These obfuscations variously aim at creating stuck-at faults, delay faults, virtual connections which are stealthy to the attacker, or to increase the reverse engineering effort by introducing extraneous information. A summary of the impacts of these device-level techniques is given at the end of this section in Tab.~\ref{tab:fault_mechanisms}. In the following we focus on the most significant ways for each mechanism to influence the behavior of the design, but note that there is significant overlap and ambiguity between them, for example, steathly signaling, and stuck-at faults or timing faults.

\subsection{Device Specific Mechanisms}
\label{sec:dev_mech}

\subsubsection{Source/Drain Doping for Stuck-at or Timing Faults}
\begin{figure}
\centering
\includegraphics[width=0.9\columnwidth]{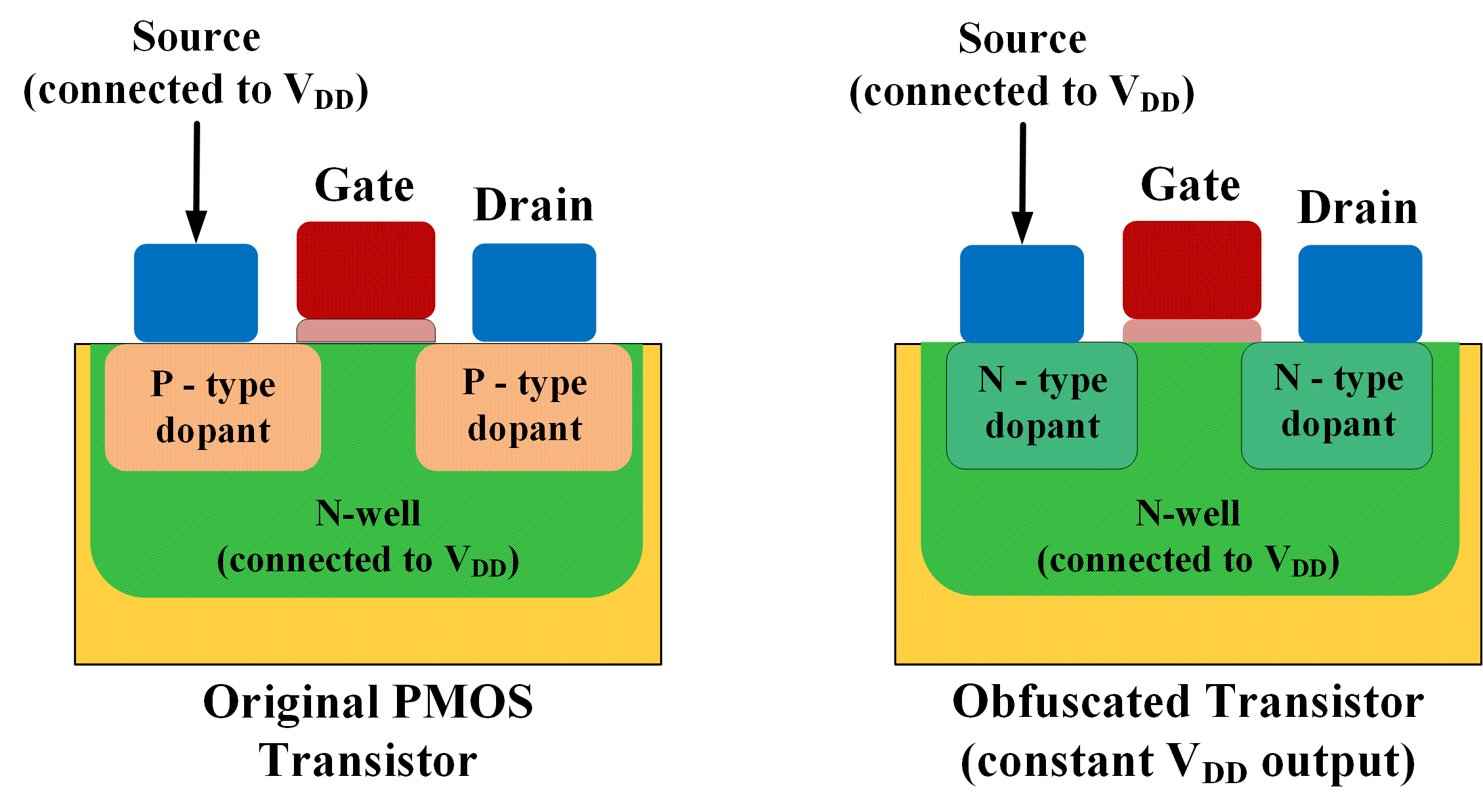}
\caption{Use of atypical doping to make an apparent PMOS transistor realize a constant-$V_{DD}$ output.}
\label{fig:dopant_manipulation}
\end{figure}

In semiconductor manufacturing doping is one of the major techniques to modify the electrical characteristics of transistors.  Doping concentration can be used to change transistor characteristics with little or no change in transistor geometry, which makes it hard for an adversary to observe them directly. \figurename~\ref{fig:dopant_manipulation} shows the layout of a PMOS transistor; under normal conditions, the source and drain regions are doped with P-type dopant. If those regions are instead doped with N-type dopant, a short circuit is created between the drain and source terminals. This type of dopant manipulation has been used to create stuck-at faults and create Trojans \cite{becker-13}. Similarly, a circuit with such dopant characteristics can be used to confuse the attacker by providing pretentious information about the circuit's functionality. An example of how this can be exploited for obfuscating CMOS circuit is discussed in Section \ref{sec:cmoscircuit}.

\subsubsection{Channel Doping for Stuck-at or Timing Faults}
Similar to source/drain doping, channel doping can be varied to change the characteristics of transistor. By controlling channel doping, a transistor can be configured as depletion or enhancement type. In an enhancement type, the channel is implanted rather than induced, the transistor is always \texttt{on}. This can be used to create stuck-at faults. An example of how this can be used for obfuscating CMOS circuit is discussed in Section \ref{sec:cmoscircuit}. 

Multi-threshold design is a well-known technique where multiple threshold transistors are used in the same digital design to achieve power-performance trade-offs \cite{kao1997transistor}. As the threshold of a transistor depends on channel doping, channel doping can be used to subtly manipulate performance to create non-obvious designs. Examples of how threshold voltage can be exploited in flip-flops, pass transistors and dynamic logic circuits are discussed in section Sections \ref{sec:ffs}, \ref{sec:passTxs} and \ref{sec:dynamicDCVS}, respectively. Multiple publications and patents deal with the usage of dopant manipulation techniques to achieve obfuscation \cite{syphermedia-library, baukus-98-patent,baukus1999digital,baukus2001secure,cocchi2012building, malik-obfusgate}. 

\subsubsection{Reversing Dopant Obfuscation}
From an attacker's perspective, even though doping does not change geometry, it is possible to detect source/drain doping changes by using passive voltage contrast (PVC) method~\cite{sugawara-14}. However, PVC requires the use of scanning electron microscope (SEM) or focused ion beam (FIB), and this makes PVC analysis slow, especially when there are millions of transistors involved. An attacker can also use picosecond imaging circuit analysis (PICA)~\cite{tsang2000pica} to detect channel doping, but this is more expensive than PVC~\cite{sugawara-14}.

\subsection{Interconnect Specific Mechanisms}
\label{sec:inter_mech}

\subsubsection{Manipulating Inter-Layer Dielectric for Timing and Stuck-at Faults} 
\label{sec:metal_fill}
\begin{figure}
\centering
\includegraphics[width=0.9\columnwidth]{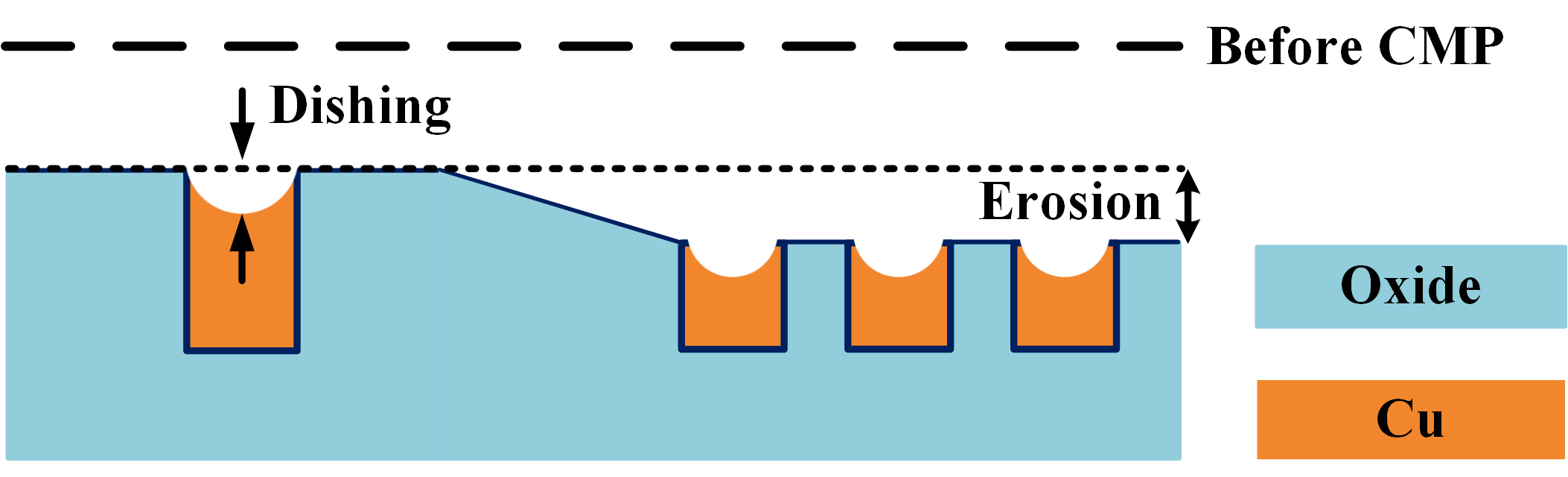}
\caption{Dishing and Erosion}
\label{fig:dishing}
\end{figure}

Chemical Mechanical Planarization (CMP) is the process used to flatten each fabricated layer of a circuit after printing the layer features. CMP can adversely affect the final layer profile due to the phenomena of \textit{dishing} and \textit{erosion} as shown in \figurename~\ref{fig:dishing} \cite{chen2010contrast}. The extent of dishing and erosion depends on line width and density~\cite{chen2010contrast}. Metal-fills help to provide line density uniformity across chip to allow for predictable CMP performance, and for this reason pattern density is commonly constrained by design rules. A basic model for the relation of ILD thickness (z) and pattern density at a given location (x, y) was proposed by Ouma~\cite{ouma1998modeling} and is given by (\ref{eqn:cmp}). Manipulating the variables of (\ref{eqn:cmp}) allow deliberate control of dishing and erosion for the purpose of obfuscation. For example, metal-fills surrounding an interconnect line can degrade the line during fabrication if placed at improper distances from it~\cite{suresh2012lithography}; in obfuscation scenarios, this can be used to create a deliberate stuck-at or timing fault.

\begin{equation}
z = \begin{cases} 
z_0 - \frac{Kt}{\rho_0(x,y)} & \text{if } t < \frac{\rho_0 z_1}{K} \\
z_0 - z_1 -Kt + \rho_0 z_1  & \text{if } t > \frac{\rho_0 z_1}{K} \\
\end{cases}
\label{eqn:cmp}
\end{equation}

where:
\begin{itemize}
\item $K$ – oxide polishing rate
\item $z_0$ – thickness of oxide deposition
\item $z_1$ – initial step height
\item $t$ – total polish time
\item $\rho_0(x, y)$ – initial oxide pattern density before CMP
\end{itemize}

\subsubsection{Dummy Logic to Increase Reverse Engineering Effort} 

In semi-custom design flows, with cell rows and standard cells, not all rows can be fully utilized by the design. The resultant unused space is filled with extra gates, called back-fills. This extra logic is useful to make late-stage design changes, for debug, or for replacing bad cells during fabrication. Obfuscation can be achieved by combining unused back-fills with metal-fills to create junk logic.  This increases complexity of the design that an attacker needs to decipher. Effective utilization of metal fills to harden a design against an attacker is proposed by Baukus \etal\ \cite{baukus2005multilayered} and Cocchi \etal\  \cite{cocchi2014cktcamo}. In addition to increasing the volume of data that an attacker needs to process, there is also increased uncertainty as to which metal traces are real and which are extraneous. Further, connecting the metal fills to power, ground, switching signals or keeping them floating reduces the effectiveness of reverse engineering techniques like voltage contrast.

\subsubsection{Stealthy Signaling using Crosstalk} 

\begin{figure}
\centering
\includegraphics[width=0.9\columnwidth]{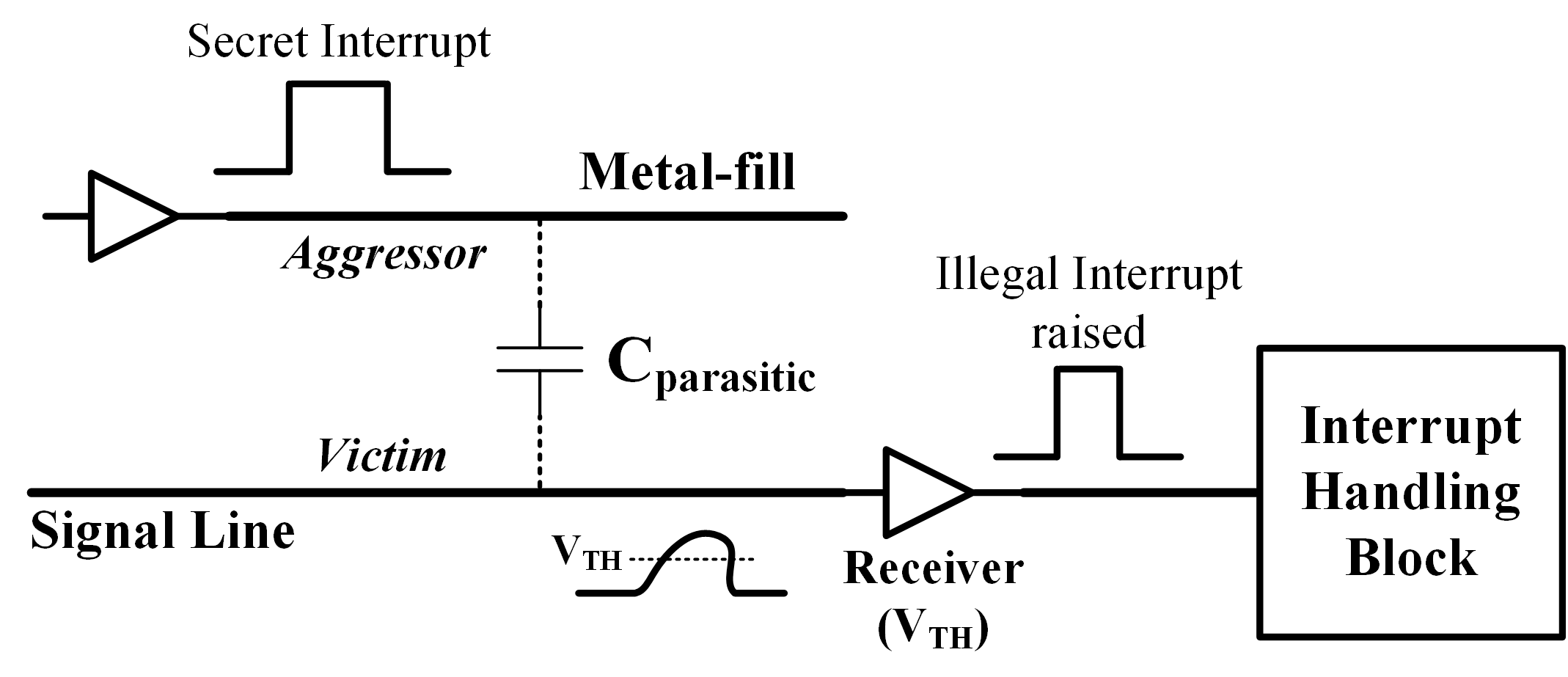}
\caption{Illustration of a Metal-fill influencing a signal line via capacitive coupling}
\label{fig:interrupt}
\end{figure}

Metal-fill can also be used for stealthy signaling by exploiting the parasitic capacitance between neighboring interconnects~\cite{shilimkar2011closed,kima2007simple}. Metal-fills are usually connected to VDD/ GND to reduce crosstalk effect on nearby signal lines. If the Metal-fills nearby to a target signal line are connected to a clock or other controllable line, then they will induce strong and controllable crosstalk on the signal line. For example, using crosstalk in this manner, the Metal-fill can raise an interrupt signal by inducing a noisy signal higher than the threshold level ($V_{TH}$) of the buffer on an interrupt line. Through proper design, only the designer has knowledge of when the interrupt will be raised. This can be utilized to obfuscate the functioning of a particular block by creating a secret interrupt unobservable by an attacker. We illustrate this in \figurename~\ref{fig:interrupt}. 

\begin{figure}
\centering
\includegraphics[width=0.9\columnwidth]{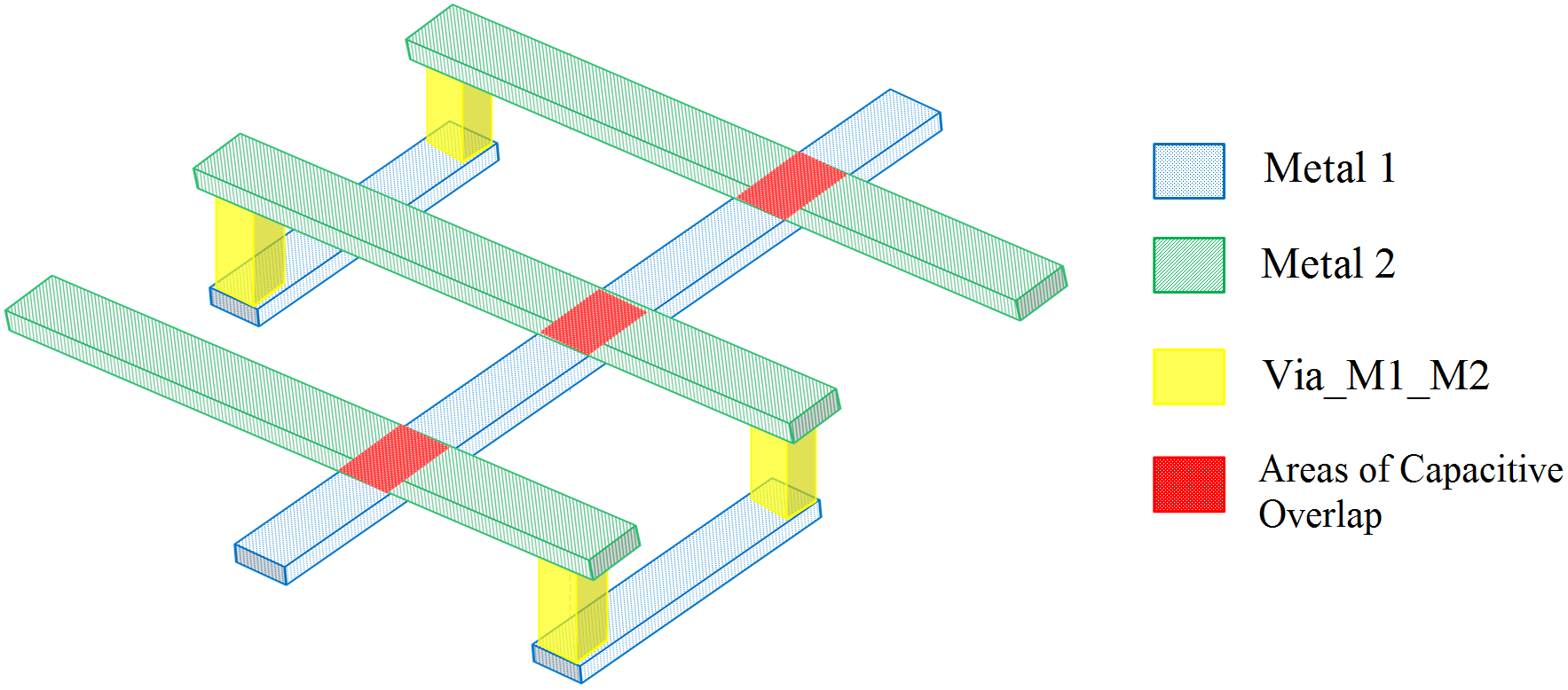}
\caption{Zigzag routing for crosstalk}
\label{fig:crosstalk_zigzag}
\end{figure}

\label{sec:ILD}
Manipulation of Inter-Layer Dielectric (ILD) is a potential control knob for facilitating stealthy capacitive signaling using crosstalk. ILD is made from low-$\kappa$ dielectric materials to enhance interconnect electrical performance by reducing capacitive coupling and the resultant crosstalk noise between adjacent lines. Air pockets are also introduced to further reduce the effective $\kappa$ but care should be taken not to compromise the mechanical strength of the ILD~\cite{raghavan2012interlayer}. To support obfuscation, a particular ILD layer can be thinner than normal or made of higher $\kappa$ material to increase capacitive coupling between interconnects across adjacent layers and increase crosstalk. Crosstalk model for two adjacent wires is shown in (\ref{eq:xtalk}) and (\ref{eq:k}). Since metal lines in adjacent layers run perpendicular to each other, one can maximize crosstalk by running one interconnect, the \textit{victim}, in one layer and another interconnect, the \textit{aggressor}, in adjacent layers in a zigzag pattern (\figurename~\ref{fig:crosstalk_zigzag}). ILD thinning maximizes the crosstalk induced by this inter-layer zig-zag pattern. Without extensive electrical measurement to estimate the magnitude of crosstalk that may occur between a pair of neighboring lines, a reverse engineer will find it difficult to extract any functionality that depends on crosstalk. 

\begin{equation}
\Delta V_{victim} = \frac{C_{adj}}{C_{gnd} -v +C_{adj}}\frac{1}{1+k} \Delta V_{aggressor}
\label{eq:xtalk}
\end{equation}

\begin{equation}
k = \frac{\tau_{aggressor}}{\tau_{victim}} = \frac{R_{aggressor}\left(C_{gnd-a} + C_{adj} \right)}{R_{victim}\left(C_{gnd-v}+C_{adj}\right)}
\label{eq:k}
\end{equation}

\subsubsection{Via Alterations for Stealthy Signaling} 

Via manipulation techniques for obfuscation rely on the fact that most reverse engineering efforts delayer the chip without detailed inspection of the vias that connect the layers to each other. An attacker usually assumes the existence of a via based on the information from the metal layers, thereby recovering incorrect logic. Chow \etal\ \cite{chow2002integrated} first proposed using a dummy via to give the appearance of a connection between a M1 layer metal and source/drain of a transistor. However, the inserted via ended in a field oxide close to the active layers of the transistor. Chow \etal\ \cite{chow2004integrated} later proposed using dummy vias between interconnects on two metal layers such that there appears to be a connection, but in reality there is not. Rajendran \etal\ \cite{rajendran-13} have showcased the use of a generic layout with multiple via sites. The vias are either designated as true or dummy during design, and the same layout is used to implement multiple functionality. 

The process of via formation has been described extensively \cite{kim2012novel,kinoshita2008via}. Controlling via formation provides another venue for obfuscation. One possible obfuscation technique is to increase ILD thickness such that the via is not fully formed or is very thin and breaks during burn-in. This can be utilized to create stuck-at faults in a circuit known only to the designer. Effective ILD thickness can be increased locally without affecting the rest of the layer as explained previously by using metal-fill density to control \textit{dishing} and \textit{erosion} as shown in \figurename~\ref{fig:dishing}~\cite{chen2010contrast}. 

\subsubsection{Lithographic Printability Features to Manipulate Timing}
\begin{figure}
\centering
\includegraphics[width=0.9\columnwidth]{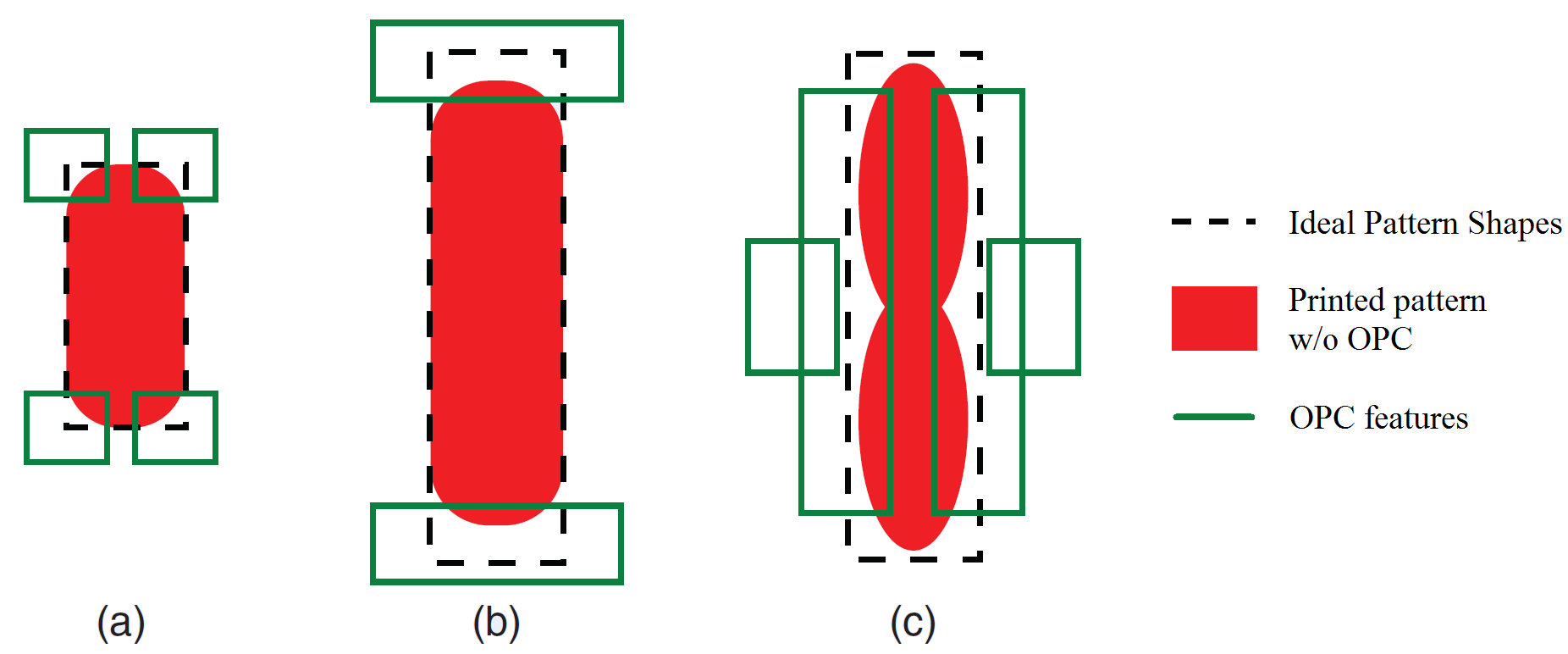}
\caption{OPC compensation techniques: (a) Serifs for Corner-rounding, (b) Hammerheads for Line-end shortening and (c) Line Biasing for Line-width shrinking}
\label{fig:opc_compensation}
\end{figure}

%\subsubsection{Interconnect Mask Manipulation}
\label{sec:IMM}
Optical Proximity Correction (OPC) and Sub-Resolution Assist Features (SRAFs) are techniques used to improve the printability of patterns on a fabricated chip. In obfuscation, these techniques can be used to give fine-grained control over fabricated structures and their delays. OPC addresses three major types of pattern distortions that occur during the fabrication process --- corner rounding, line-end shortening and line-shrinking. These are compensated for by use of \textit{serifs}, \textit{hammerheads} and \textit{line-biasing} OPCs, respectively~\cite{chen2009routing,sameer2004interconnect}. The various distortions and their associated compensations are illustrated in \figurename~\ref{fig:opc_compensation} \cite{sameer2004interconnect}.

In obfuscation, changing or removing \textit{line-biasing} compensation enables thinning of the interconnect and increasing the line resistance. This affects the delay through the interconnect due to changes in the \textit{RC} time constant, where \textit{R} is the line resistance and \textit{C} is the capacitive load on the line. The delay change can be utilized by the designer to introduce delay faults between flip-flops where a particular signal is biased such that it always arrives late and is never latched to the next stage. If a reverse engineer is unaware of this, they would infer incorrect logic from this line.

\begin{figure}
\centering
\includegraphics[height=1.5in, width=0.7\columnwidth]{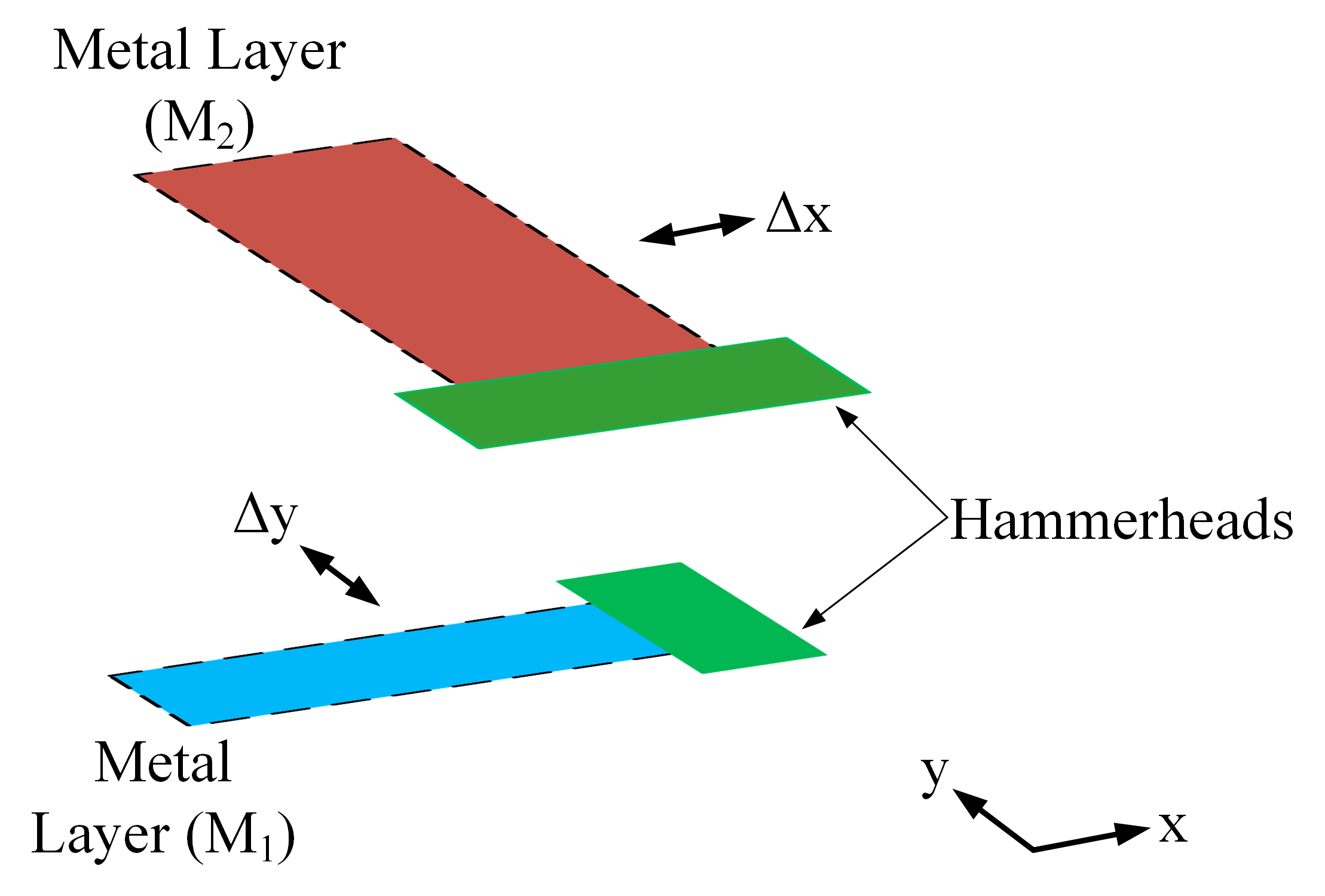}
\caption{Interconnect mask lines with Hammerheads in two adjacent layers}
\label{fig:inter_mask2}
\end{figure}

Manipulation of \textit{hammerheads} can similarly slow down signals when they cross layers through vias at the end of the lines. Size of the \textit{hammerheads} account for the effects of lithography, etching and also, for mask misalignment tolerances. By shaping or removing the \textit{hammerheads} and changing the line alignments we can create an open on the interconnect crossing layers. Also, we can create a weak connection between two lines on different layers by partial misalignment, as illustrated in \figurename~\ref{fig:inter_mask2} where $\Delta x$ and $\Delta y$ alignments control whether the lines are connected or not.

SRAFs are features smaller than the smallest required feature printed on a die. They are used to improve the printability of the desired patterns during the lithography process, equalize lithographic performance between isolated and densely placed features, and increase wafer yield. Melvin \etal\ \cite{melvin2006assist} explain where SRAFs are usually placed. Many works~\cite{dhumane2011lithography,gupta2007detailed,shi2002understanding} also deal with methodology for efficient placement, shaping and sizing of SRAFs. While proper SRAF placement improves printability of the target feature, primarily, via reduction in edge placement error~\cite{dhumane2011lithography}, it is also possible to degrade the feature by altering the placement~\cite{shi2002understanding}. Thinning the line using SRAFs can be utilized in the same manner as \textit{line-biasing} OPC to change resistance and cause timing faults. Ideally SRAFs should not be printed. In reality, there are tiny artifacts that are left behind due to SRAFs after completion of the manufacturing process. It is possible to size the SRAFs such that these final artifacts are of significant size to affect the legal features adjacent to them, like adding parasitic capacitance to a nearby interconnect.

OPC and SRAF-based manipulations will require high resource investment to implement. Many factors that might affect the printing of features during manufacturing are difficult to control. SRAFs increase the number of constraints in the design rules and their properties are adjusted constantly during post-layout lithographic simulations to improve yield. Hence, a designer will need to work with the foundry engineers to effectively utilize OPC and SRAFs for obfuscation. However, the high expense may have a corresponding high payoff, as the small manipulations caused by OPC and SRAFs will also be very difficult for an attacker to account for in his models.

\begin{table*}[th]
  \renewcommand{\arraystretch}{1.4}    
  \setlength{\tabcolsep}{4pt}
  \centering
  \caption{Summary of fault mechanisms and corresponding obfuscation classes and detection schemes}
  \label{tab:fault_mechanisms}
  \begin{tabular}{| c | c | c | c | c | c | c |}
\cline{4-7}
 \multicolumn{3}{c}{} & \multicolumn{3}{|c|}{\textbf{Obfuscation Classes}} & \multirow{2}{*}{\parbox[c][10ex]{0.1\columnwidth}{\centering \textbf{Detection}}}\\ \cline{4-6}
 \multicolumn{3}{c|}{} & \parbox[c][7ex]{0.1\columnwidth}{\centering Stuck-at Faults} & \parbox[c][7ex]{0.1\columnwidth}{\centering Timing Faults} & \parbox[c][7ex]{0.11\columnwidth}{\centering Stealthy Signaling} & \\ \cline{1-7}
 \multirow{8}{*}{\parbox{1.5cm}{\textbf{Physical Design Mechanisms}}} &  \multirow{2}{*}{Doping} & Source/Drain & \xmark & & & PVC\\ \cline{3-7}
& & Channel & \xmark & \xmark & & PICA\\
\cline{2-7}
& \multicolumn{2}{|c|}{Metal-fill}  & \xmark & \xmark & \xmark & SEM\\ \cline{2-7}
&  \multirow{2}{*}{ILD Manip} & Thinning &  & \xmark & \xmark & -\\ \cline{3-7}
& & Thickening & \xmark &  \xmark & \xmark & -\\ \cline{2-7}
& \multicolumn{2}{|c|}{Intercon mask manipulation}  & \xmark & \xmark & \xmark & SEM, SQUID\\ \cline{2-7}
& \multicolumn{2}{|c|}{SRAFs}  &  & \xmark & & -\\ \hline
  \end{tabular}
  
\end{table*}
\subsubsection{Reversing of Interconnect Manipulations}

For the techniques that use mask manipulations or metal-fills, an attacker using imaging techniques, like SEM, can only observe features in two dimensions (x,y). The delayering process removes the vertical dimension information. The attacker will require SEM to study the interconnect thickness variations to comment on changes in time constants. The interconnect obfuscation mechanisms are more useful in the lower metal layers where the majority of the signals are routed and the interconnect density is high. An attacker will have to invest a large amount of time to find potential sites of interest. 

Also, interconnect misalignment techniques may not be detected due to the requirement of correlating measurements in two different lower metal layers which may fall below the resolution of the microscopy technique used.
It is possible for an attacker to detect open/short faults on interconnects of interest using Superconducting Quantum Interference Device (SQUID) microscopy~\cite{nikawa2001squid}. The attacker will still need to narrow the search to the target area by using other electrical techniques.
To get information about the dielectric thickness of various layers, an attacker will need to slice the chip and observe the vertical layers through some microscopy technique. However, the dishing and erosion effects are localized and may not be reflected at the plane of the cut. Hence, inter-layer dielectric manipulation to affect crosstalk and via alterations cannot be easily found through microscopy techniques.

\section{Logic-Level Mechanisms}
\label{sec:logical_mechanisms}
The second aspect of physical design obfuscation is concerned with logic-level structures. Circuit structures are needed to translate the low-level device and interconnect obfuscation mechanisms of the previous section into building blocks, which can be used for realizing complex sequential logic functions. These (primarily combinational) circuit structures form the next level of abstraction. As mentioned earlier, they can sometimes be agnostic of the underlying physical mechanisms (i.e. sizing, dopant, etc). In this section, we discuss novel logic-level techniques for obfuscation based on device-level mechanisms described in Section V.

\subsection{Circuit Transformations for Obfuscation}
\label{sec:ckt_tech}
The device-level mechanisms described in Section \ref{sec:physical_mechanisms} have to be coupled with circuit topologies to create obfuscated circuits. The physical mechanisms can, for example, create a stuck-at faults or timing faults in a circuit to change its behavior without changing its appearance. After employing these mechanisms to change the circuit behavior, the realized function of the circuit will differ from its apparent function. Because the modified circuit may violate common circuit truths, e.g., signal propagation and timing, a reverse engineer cannot successfully assume these truths to hold when trying to reconstruct the circuit function.  If a reverse engineer can no longer exploit these simplifying assumptions, then his task will become significantly more challenging. In the following subsections, we present examples to show that different circuit implementation styles present different opportunities for obfuscation.

\subsubsection{CMOS circuits}
\label{sec:cmoscircuit}
\begin{figure}
\centering
\includegraphics[height=1.5in, width=0.6\columnwidth]{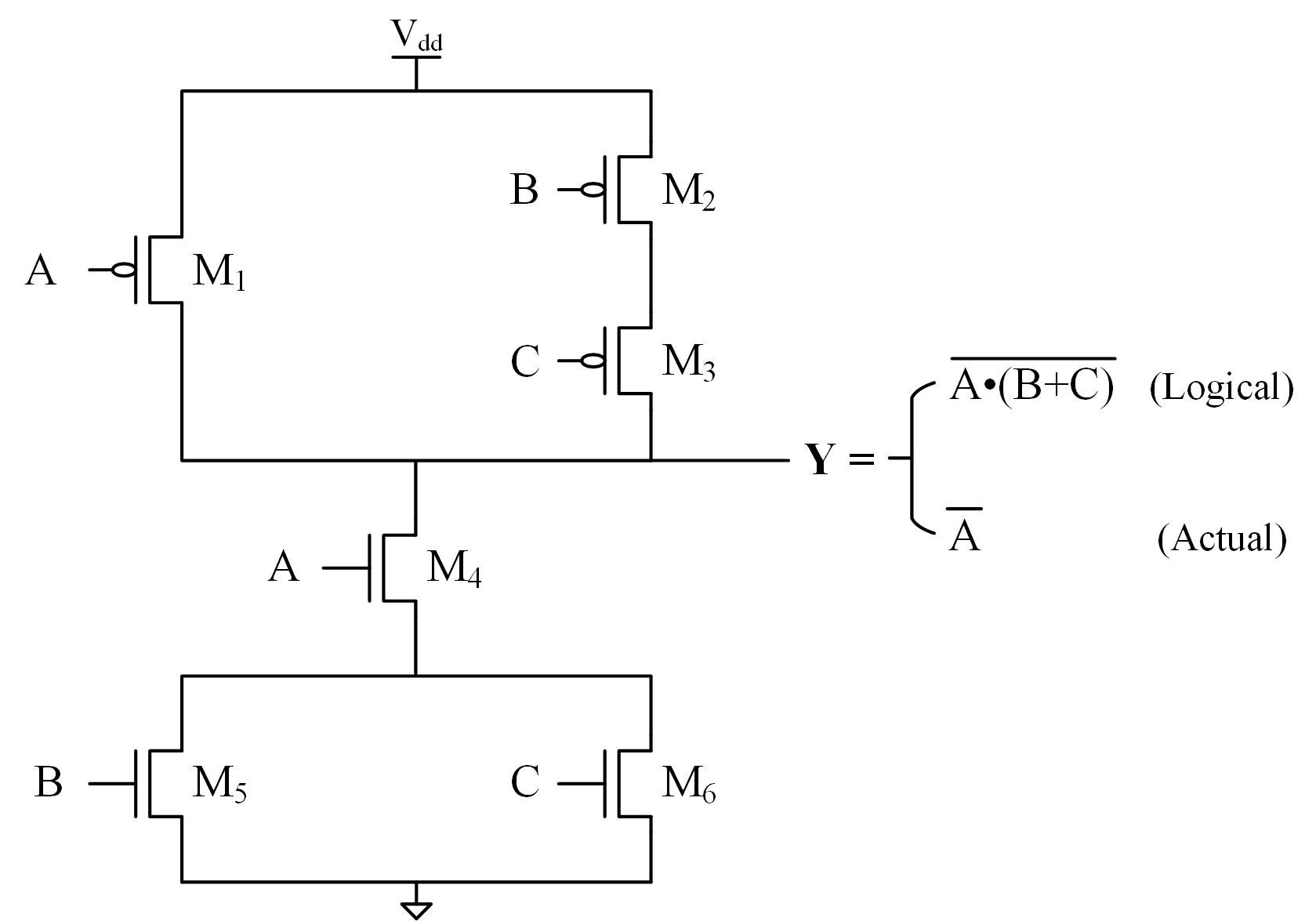}
\caption{An example of CMOS logic obfuscation}
\label{fig:cmos}
\end{figure}

CMOS circuits can be obfuscated to protect the design and confuse an attacker by using various device manipulations. For example, consider the circuit shown in \figurename~\ref{fig:cmos} implementing the logic $Y=\overline{A\cdot(B+C)}$. Using the mechanisms described in Section \ref{sec:dev_mech} the circuit can be morphed to realize a logic different than its apparent one. If transistor $M_{5}$ and $M_{6}$ are converted from enhancement type to depletion type MOSFETs by doping, they will be always \texttt{on}. Similarly, if transistor $M_{2}$ is converted to an open circuit using source/drain doping, the node between $M_{2}$ and $M_{3}$ will be floating~\cite{becker-13}. These changes would together cause the circuit to realize the functionality of an inverter $Y=\bar{A}$ instead of its apparent function of $Y=\overline{A\cdot(B+C)}$. Similarly, the circuit function could be modified by using the mechanisms described in Section \ref{sec:inter_mech}, such as via alteration, to change the behavior of the interconnects that drive the circuit inputs. For example, if \textit{SA1} faults are created on the interconnects driving the $B$ and $C$ inputs, the circuit would again realize the function of an inverter, even if the transistors are unmodified.

\subsubsection{Pass Transistors}
\label{sec:passTxs}
Pass transistors are used as logic switches to create paths between the input and output nodes. Relative to CMOS logic, pass transistor logic is bidirectional and can be faster and reduce transistor count and power. Hence, pass transistor logic is often used in multiplexers, switches and efficient XOR implementations. The doping mechanisms from Section \ref{sec:dev_mech} allow for controlling the threshold voltage of pass transistors to create chosen stuck-at faults that modify circuit behavior. 

\begin{figure}
\centering
\includegraphics[height=2in, width=0.7\columnwidth]{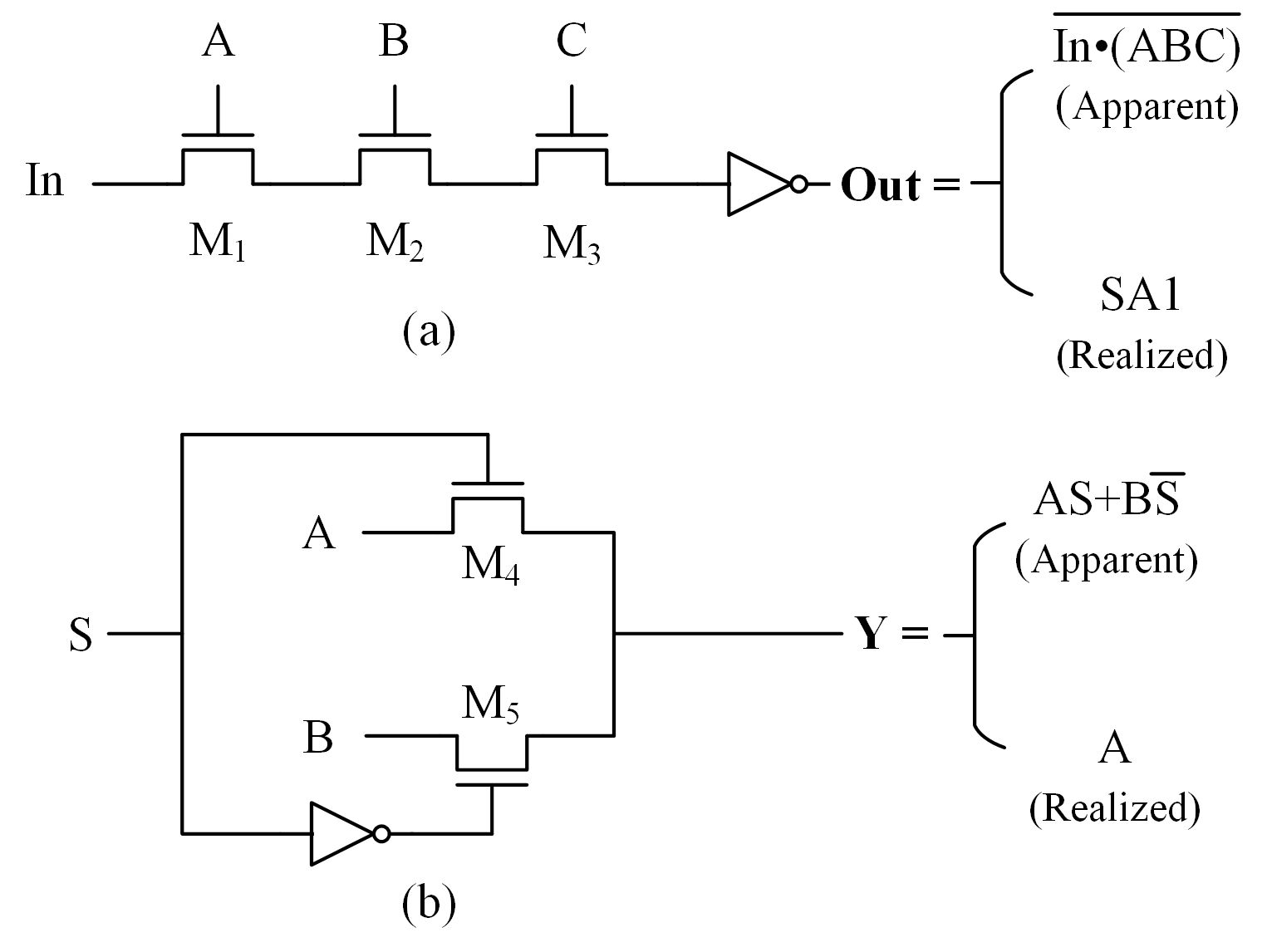}
\caption{Pass transistor circuits. (a) Cascaded pass transistors, (b) 2:1 Mux using pass transistors}
\label{fig:pass_transistors}
\end{figure}

\figurename~\ref{fig:pass_transistors} (a) shows cascaded logic using three pass transistors. It is possible to obfuscate the logic by using a high-threshold transistor for $M_{1}$ so that the voltage drop through the transistor is large enough for the cascade output to always be perceived as logic-0. This means that the input to the inverter at the end of the cascaded logic is always logic-0 and node $Out$ realizes a stuck-at logic-1. The values applied to the gates of pass transistors $M_2$ and $M_3$ are irrelevant in this example, but this would not be apparent to the attacker.

\figurename~\ref{fig:pass_transistors} (b) shows an apparent $2:1$ multiplexer where one can similarly induce faults such that $M_5$ is always open and $M_4$ is always closed. The realized logic function after these changes would be $Y=A$ instead of a multiplexer.

\subsubsection{Dynamic Logic and Differential Cascode Voltage Switch (DCVS) Logic}
\label{sec:dynamicDCVS}
\begin{figure}
\centering
\includegraphics[height=1.5in, width=0.75\columnwidth]{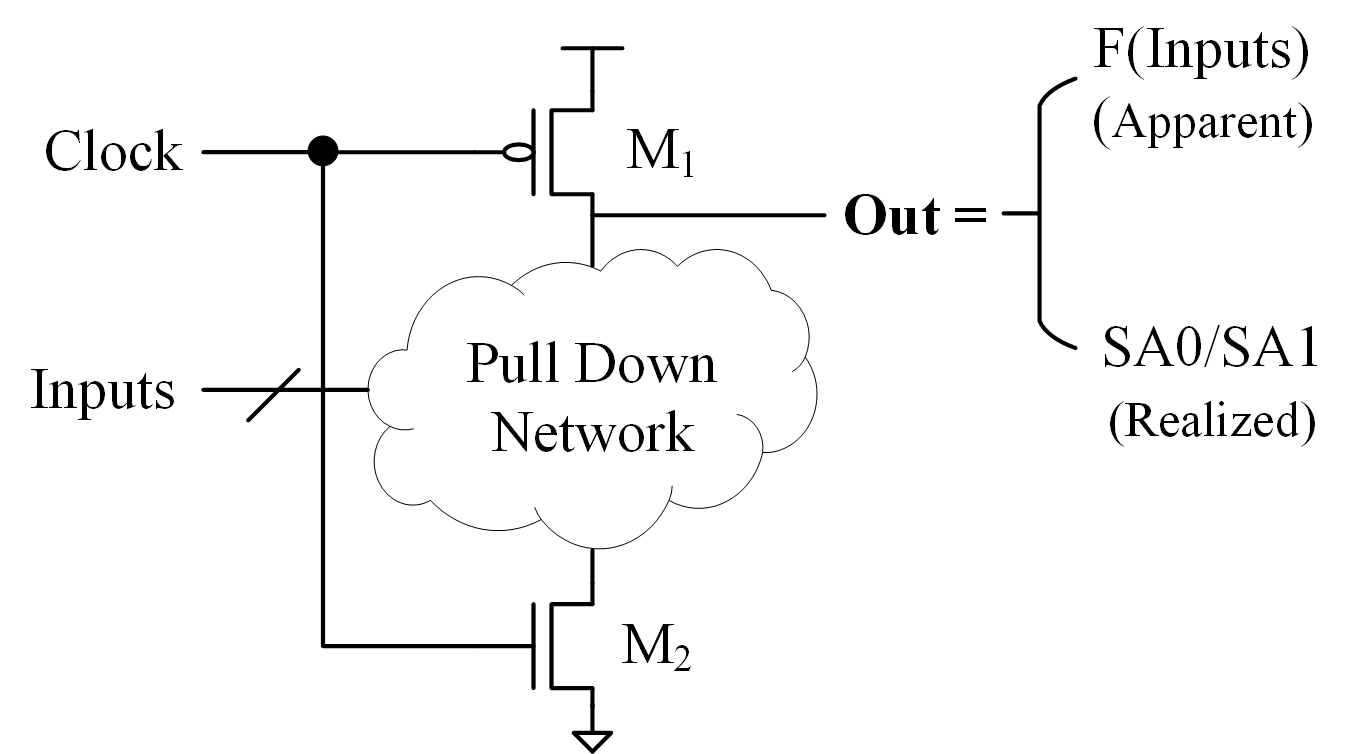}
\caption{Dynamic Logic}
\label{fig:dynamic_logic}
\end{figure}

\begin{figure}
\centering
\includegraphics[width=0.8\columnwidth]{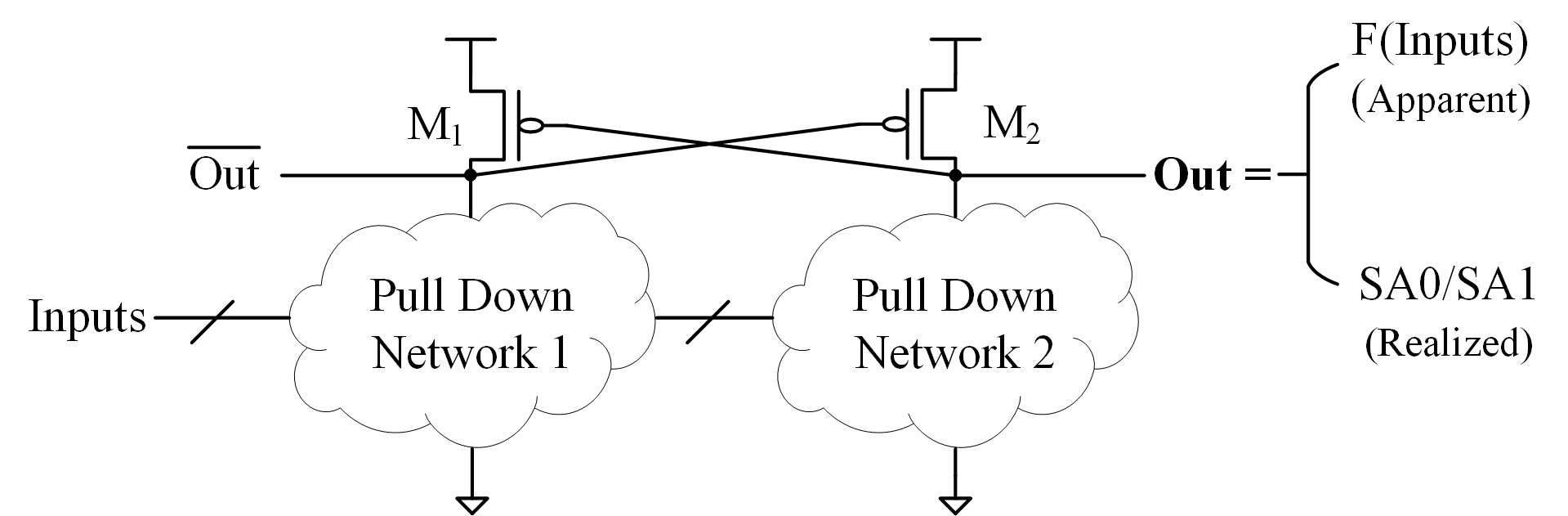}
\caption{Differential Cascode Voltage Switch (DCVS) Logic}
\label{fig:dvcs}
\end{figure}

Dynamic logic and Differential Cascode Voltage Switch (DCVS) logic finds use in high-speed applications \cite{somasekhar1996differential} where an NMOS pull-down network is preferable for performance reasons. \figurename~\ref{fig:dynamic_logic} and \figurename~\ref{fig:dvcs} illustrate dynamic logic and DCVS logic, respectively. The two circuits as depicted are sensitive to changes in transistors $M_{1}$ and $M_{2}$, so a natural approach to obfuscation is to manipulate these transistors. For example these transistors can be made stronger or weaker using doping and channel length biasing as described in Section \ref{sec:dev_mech}, creating stuck-at faults at the outputs irrespective of the inputs. An attacker that cannot observe the properties of $M_{1}$ and $M_{2}$ would not be able to discern that the computational logic is irrelevant to the output value. 

\subsubsection{Flip-flops}
\label{sec:ffs}
\begin{figure}
\centering
\includegraphics[width=0.9\columnwidth]{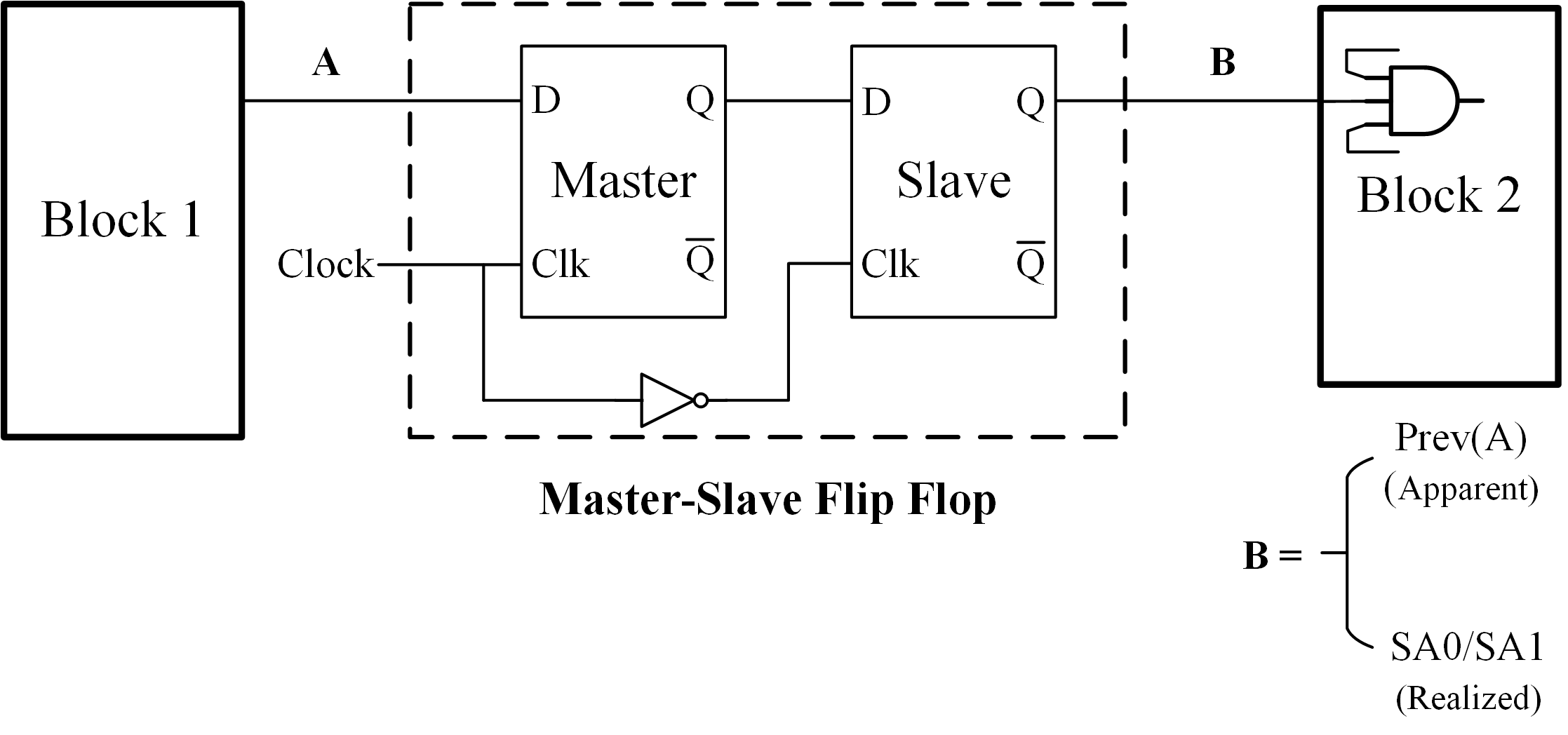}
\caption{Flip-flops under obfuscation}
\label{fig:flip_flops}
\end{figure}

Large sequential digital systems are composed of combinational logic and state-holding elements such as flip-flops. A flip-flop circuit that may appear to be a state-holding element can be manipulated to instead always hold a particular value. For example, consider the circuit shown in \figurename~\ref{fig:flip_flops}. Signal $A$ is connected as the input to the flip-flop and from appearances would be expected to be sampled on the clock edge and passed on to the output of the flip-flop. In reality, by inducing a stealthy stuck-at fault, the output of the flip-flop can be assigned logic-0 or 1. For an attacker who tries to extract the logic, the circuit may appear to have a logical connection from Block 1 to Block 2 but the realized function has no such connection. Multiple such scenarios can be created with flip-flops that are inserted only for the purpose of increasing the complexity of reverse engineering the design. 

\begin{figure}
\centering
\includegraphics[width=0.9\columnwidth]{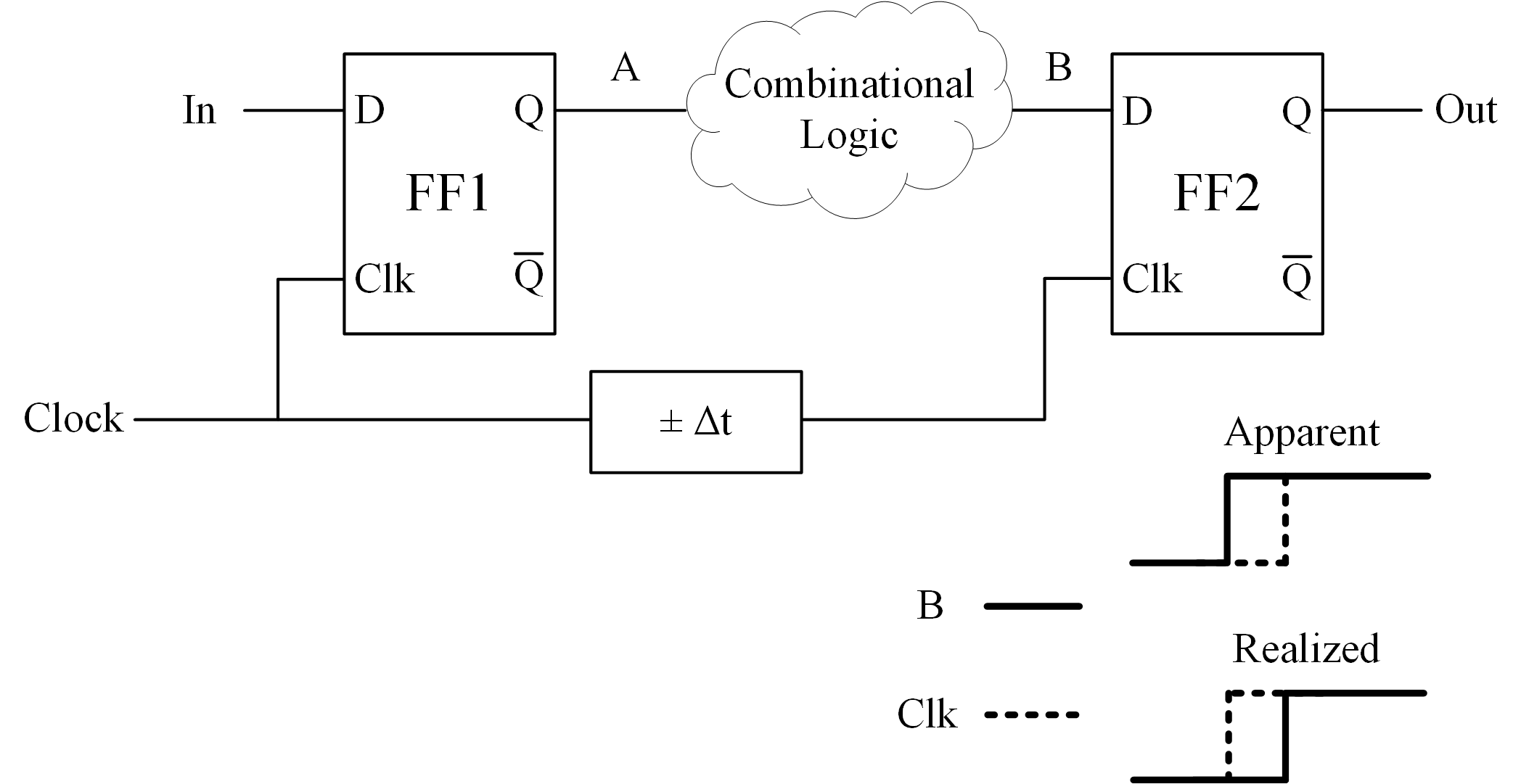}
\caption{Setup and Hold Violations illustration}
\label{fig:setup_hold}
\end{figure}

{\itshape Setup and Hold Violations:}
Synchronous sequential systems with flip-flops typically need to obey setup and hold time constraints to function as intended. Setup time requirements specify that flip-flop input signals must not change immediately before a clock edge, and hold time requirements specify that flip-flop input signals must not change immediately after a clock edge. Intentionally violating these requirements can create predictable errors that can be employed for obfuscation. For example, consider the circuit shown in \figurename~\ref{fig:setup_hold}. The path from $FF_{1}$ to $FF_{2}$ through the combinational logic has to obey the setup time requirements at $FF_2$ for a given clock period to operate without error, and a reverse engineer would typically assume this to be the case. If the devices and interconnects on a logic path were slowed down using techniques such as high-threshold transistors and interconnect thinning (Sec.~\ref{sec:physical_mechanisms}), the setup time would be violated, and the realized sequential behavior would differ from the apparent behavior. Similar to setup time violations, hold time violations can be induced but may require compound manipulations to speed up paths and induce errors. 

\begin{figure}
\centering
\includegraphics[width=0.9\columnwidth]{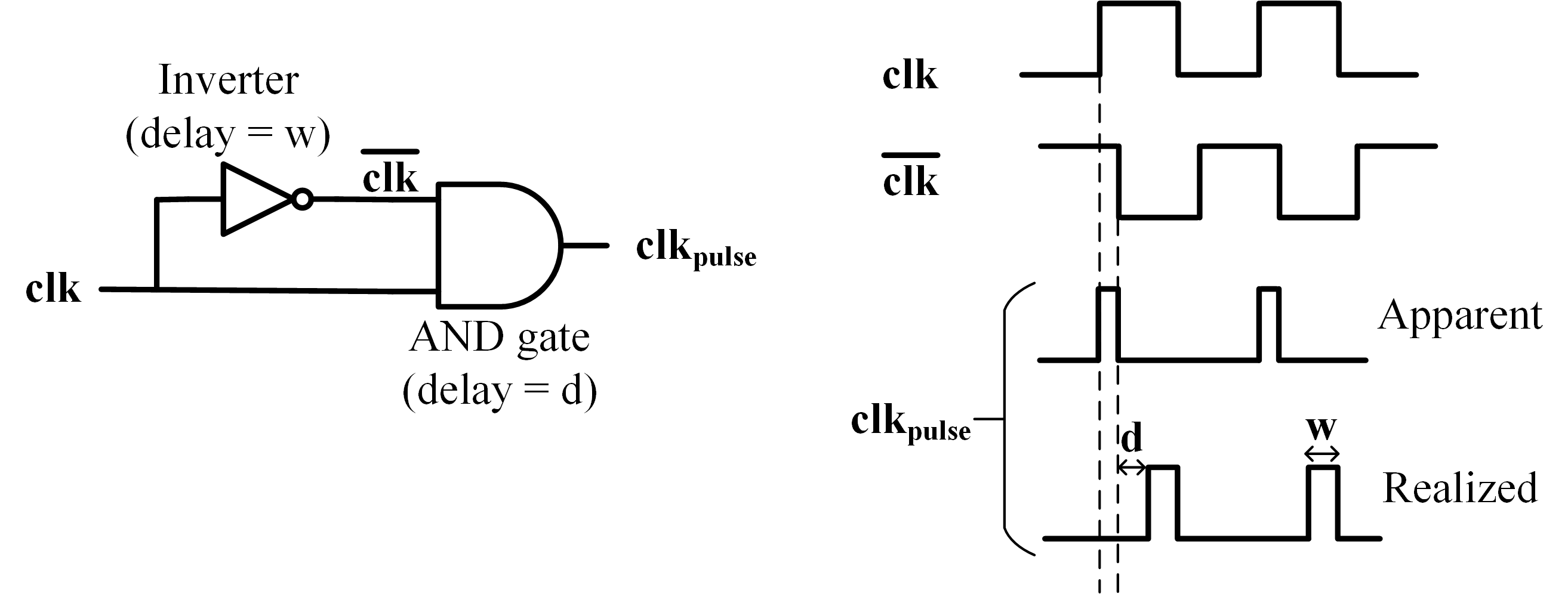}
\caption{Manipulating the clock signal used in Pulsed Latch }
\label{fig:pulsed}
\end{figure}

{\itshape Pulsed Latch:}
Pulsed latch based flip-flop designs are common in high-performance systems where they are deployed to improve performance and reduce power dissipation \cite{shin2011pulsed}. Pulsed latches retain the advantage of both flip-flop and latch-based design. Instead of using master-slave latches with a clock, the pulsed latch has a single latch and uses a pulsed clock so that the latch is only open for a short time. The circuit to generate a pulsed clock, and the associated waveforms, are shown in \figurename~\ref{fig:pulsed}. The behavior of the latch critically depends on the pulse width and timing, and one can subtly induce setup and hold time violations or race-through conditions by manipulating the pulsed clock. The width and timing of the pulsed clock can be manipulated by changing the delay of the inverter and the AND gate, respectively; slowing down the inverter increases the pulse width, and slowing down the AND gate will delay the arrival of the pulsed clock. An attacker would typically assume that a pulsed latch would be free of conditions such as flow-through, so these changes would cause the realized function of the latch to differ from the apparent one. Timing manipulations may be challenging for an attacker to detect during scan testing, because the scan chain is usually run at much slower speeds than the standard circuit operation.

\subsection{Promoting Physical Mechanism to Logic Level}

The previous subsection dealt with numerous ways to manipulate circuits, but now we focus specifically on gates that use a set of indistinguishable physical structures to implement different Boolean functions. An adversary observing this ambiguous physical structure will not know which logical variant of the gate he is seeing. Yet, it is important to note that, when using indistinguishable gate structures to realize multiple functions, the obfuscated gates will be larger in size and have a different pattern than the non-obfuscated gate variants that implement the same functions. So, if a design uses a mixture of obfuscated and non-obfuscated gates, then the attacker will likely have knowledge of which gates are obfuscated, which can help localize the uncertainty for reverse engineering. The reverse engineering problem becomes even more constrained when only a small number of gates are obfuscated~\cite{rajendran-13} in this way. By contrast, for the mechanisms in the previous subsection, an attacker may have to consider every component in the design as potentially misleading.

\subsubsection{Look-up tables}
Any physical mechanism that creates obfuscated 0 or 1 constants can be used to program obfuscated look-up tables for fully programmable logic gates. An $n$-input look-up table requires $2^n$ programming bits to store the desired output bit for every input combination. The programming bits would be implemented as nodes that are forced to be either stuck-at 1 or stuck-at 0 using one of the techniques from Sec.~\ref{sec:physical_mechanisms}. For example, the programming bit could have vias connecting to ground and vdd, and the attacker would have to determine which via is the dummy and which is real to know whether the corresponding gate output value was 0 or 1. An advantage of look-up tables is that they are fully general logic structures, but the disadvantage is that they are much larger than the CMOS logic gates that perform the same functions.

\subsubsection{Programmable Logic Array (PLA) Cross Points}

As with look-up tables, obfuscated connections can be used to program a PLA to resist reverse engineering. The PLA cross points that are connected would be indistinguishable from the open connections from the perspective of attacker. 

\subsubsection{Restricted Choice of Gate Functions}

A less general alternative to look-up tables is to use gates that can implement just a few different logical functions, with the choice among the functions being determined by some hard to observe mechanism such as doped channel stops~\cite{baukus-98-patent}. The work of Rajendran \etal~\cite{rajendran-13} addresses the problem of how to use a small number of obfuscated gates to efficiently obfuscate an overall circuit function. Their work specifically considers a 2-input gate structure that has indistinguishable variants for implementing XOR, NAND, or NOR functions; learning which function the gate implements requires ability to distinguish between true vias and dummy vias. A similar style of obfuscated gate is proposed by Malik \etal~\cite{malik-obfusgate}.

\subsection{Obfuscation Countermeasures: SAT Solving to Assist Reverse Engineering}

Given a set of possible component implementations, and ability to query the obfuscated circuit to learn the correct computed values for any input, one can use an oracle-guided synthesis approach~\cite{jha2010oracle} to reconstruct the functionality of an obfuscated block of combinational logic. In practice, this procedure is quite effective due to the impressive capabilities of modern SAT solvers. SAT solving is used both for synthesizing new inputs that should be applied to the oracle, and is also used for deobfuscating the design once a sufficient set of input-output examples is obtained. For more details on this approach, we point interested readers to the work El Massad \etal~\cite{elmassad-15}; their work shows that the logic function of a circuit with 200 obfuscated gates, each capable of implementing NAND/NOR/XOR ($3^{200}$ possible logic functions), can be identified in less than an hour after observing just tens of input/output vectors~\cite{elmassad-15}. Later work extends El Massad's attack using incremental SAT solvers for more efficient implementation~\cite{liu-16-oracle-guided}. This sort of attack can be applied to any context where the attacker knows a set of possible functions for each component, and can apply inputs and observe corresponding outputs through a scan chain. In certain types of obfuscated circuits, there is no attempt to hide the locations of components with obfuscated functions, and only the specific functions implemented by those components are secret. The ability to clearly identify obfuscated and non-obfuscated components helps to minimize the exponential increase in the space of possible circuit functions that must be implicitly searched during oracle-guided synthesis. We conjecture that low-level stealthy changes will be less susceptible to oracle-guided synthesis approaches because an attacker will not be able to identify \textit{a priori} the non-obfuscated components in the circuit, and must therefore consider every component to be suspect yielding a much larger space of possible circuit functions that could be implemented.

\section{Discussion and Challenges}
\label{sec:discussion}

In above sections, various physical mechanisms that can be used for obfuscation were introduced. Below, we briefly discuss two related concepts,  hardware Trojans and tamper protection techniques. We conclude this paper with a discussion of challenges and open research problems. 

\subsection{Hardware Trojans vs Obfuscation}
While physical design obfuscation aims at \textit{protecting} a design through various low level manipulations as discussed above, hardware Trojans  can potentially exploit the same techniques to insert \textit{malicious} functionality into a target circuit~\cite{becker-13}. More concretely, what are sometimes called parametric Trojans~\cite{Tehranipoor:2010:SHT:1724965.1725002} are quite similar to the mechanisms discussed in Secs.~\ref{sec:physical_mechanisms} and \ref{sec:logical_mechanisms}. There are, however, several important differences between Trojans and obfuscation. First,  Trojans have malicious intent and are therefore inserted by adversaries in the design flow, while obfuscation is introduced by designers aiming to protect a circuit. Second, Trojans and obfuscation have vastly different reliability requirements. An obfuscated design is expected to work with high reliability, similar to any other circuit, or else the function implemented would be incorrect. In contrast, a low-reliability Trojan can still compromise the security of a system, even though its malicious intent, e.g., leaking of a cryptographic key, is not fully reliable. Third, often it must be possible to trigger a Trojan, e.g., through certain input values, a requirement which is entirely absent in the case of obfuscation. Finally, the requirement of being stealthy is often more crucial in the case of obfuscation, as the adversary specifically tries to overcome the stealthiness, something that is not always the case for hardware Trojans. 

\subsection{Tampering Protection}
Hardware reverse engineering usually requires physically tampering with the target circuit. 
Obfuscation techniques, such as the ones described in this contribution, are one approach to provide on-chip protection by preventing the attacker from learning the logic function. Orthogonal to obfuscation techniques are tamper protections that prevent an attacker from even accessing data or silicon layers. Various countermeasures have been proposed to protect against physical access of circuits, and we mention here only a few of them. A prominent example is the IBM 4748 co-processor. It envelopes the hardware in a tamper-sensing and tamper-reactive shell along with tamper-triggers that can signal the processor to wipe all sensitive data and burn fuses~\cite{lindemann2001building}. While this can protect cryptographic keys, it does not prevent an attacker from accessing the circuit to extract the IP. An approach to protecting the actual circuit is described in Mikulec \etal~\cite{mikulec2002explosive}, based on  a process that embeds gadolinium nitrate channels in silicon. If this layer were to be ground, it would explode and destroy nearby components. The assumption is that such a layer is placed at the top of the chip to provide protection against delayering. 

A principle problem of physical tamper protection is that an attacker often has access to more than one chip. This allows him to  devise strategies by running experiments which attempt to overcome the countermeasures. Even if several chips are destroyed, there is the risk that she eventually succceeds. The successful attack against a Trusted Platform Module (TPM) chip with strong tamper resistant features is an instructive example~\cite{tarnovsky2010deconstructing}.
Nevertheless, for  devices with low- to medium-security requirements, physical tamper protection can be an attractive approach. We note that tamper protection can be combined with the physical design obfuscation described in the contribution at hand.

\subsection{Challenges}

In this paper, we attempt to provide a systematic way to look at physical design obfuscation, together with several concrete examples for achieving low-level hardware obfuscation. As mentioned in the introduction, circuit obfuscation has only received scant attention in the scientific community and there are several challenging (and interesting) open research questions.

In the field of software, obfuscation techniques have been studied at depth and several complexity metrics have been developed. These metrics relate to \textit{potency}, \textit{stealth}, \textit{performance} and resource \textit{cost}. A good description of these can be found in Collberg \etal~ \cite{collberg1997taxonomy}.  For hardware obfuscation, there are no metrics to quantify how well a circuit has been obfuscated. Often an adversary can ``guess'' or make conjecture about the functionality of a circuit. Such guesses may be based on design size, design density, power dissipation or any number of circuit characteristics that often characterize a specific function. Availability of diagnostic methods and commercial tools can lead an attacker to find the locations where a circuit may have been obfuscated. Availability of formal methods and tools may even allow an attacker to validate her hypothesis using measurements from the circuit. Most formal methods perform search in an exponential space. However, knowing that an adversary must search an exponential space to find the correct model says little about the practical hardness of his task. Current generation of Boolean SAT solvers demonstrate impressive capability in tackling large circuits. Thus, how well a circuit has been obfuscated is a complicated question that ensnares several different research questions into one. Quantifying the \textit{hardness} of any obfuscation solution can go a long way towards engineering adoption.

Another important question is the \textit{stealthiness} of the obfuscation, i.e., the device and logic-level techniques discussed in this paper. The strength of the methods heavily depends on the capabilities of the attackers: This is difficult to quantify yet important for determining the strength of an obfuscation techniques. 
There is a large design space for combinational and sequential circuits that lend themselves to the obfuscation methods discussed in this contribution. Little is known about optimum ways to design such circuits, which provide high security and at the same time keep the area and power overhead low. Related to this question is the CAD integration of obfuscation, i.e., an automated design flow which maps an arbitrary high-level design description to a circuit with strong reverse engineering resistance. 

Another research direction is related to the practicality of any obfuscation solution. What is the impact on design cost (also known as non-recurrent engineering cost), schedule,  manufacturing overhead, performance, area, and impact on yield? Practicality of the solutions also relates to ease of integration with existing tools and methodologies, or clear description of requirements to enable development of CAD tools for automatic obfuscation. It can be argued that this practical question is crucial for the use of physical design obfuscation in practice.

\bibliographystyle{IEEEtran}
\bibliography{obfuscation}

\begin{IEEEbiography}[{\includegraphics[width=1in,height=1.25in,clip,keepaspectratio]{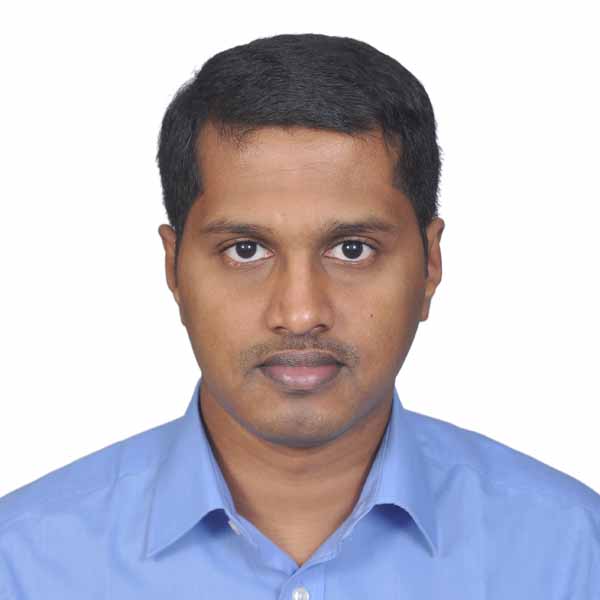}}]{Arunkumar Vijayakumar}(Student Member, IEEE)
(M\rq11) is currently pursuing his Ph.D. degree under Prof. Sandip Kundu in the Department of Electrical and Computer Engineering from the University of Massachusetts Amherst, Amherst, MA, USA. His research interests include VLSI CAD, VLSI circuit design and Hardware Security.
\end{IEEEbiography}

\begin{IEEEbiography}[{\includegraphics[width=1in,height=1.25in,clip,keepaspectratio]{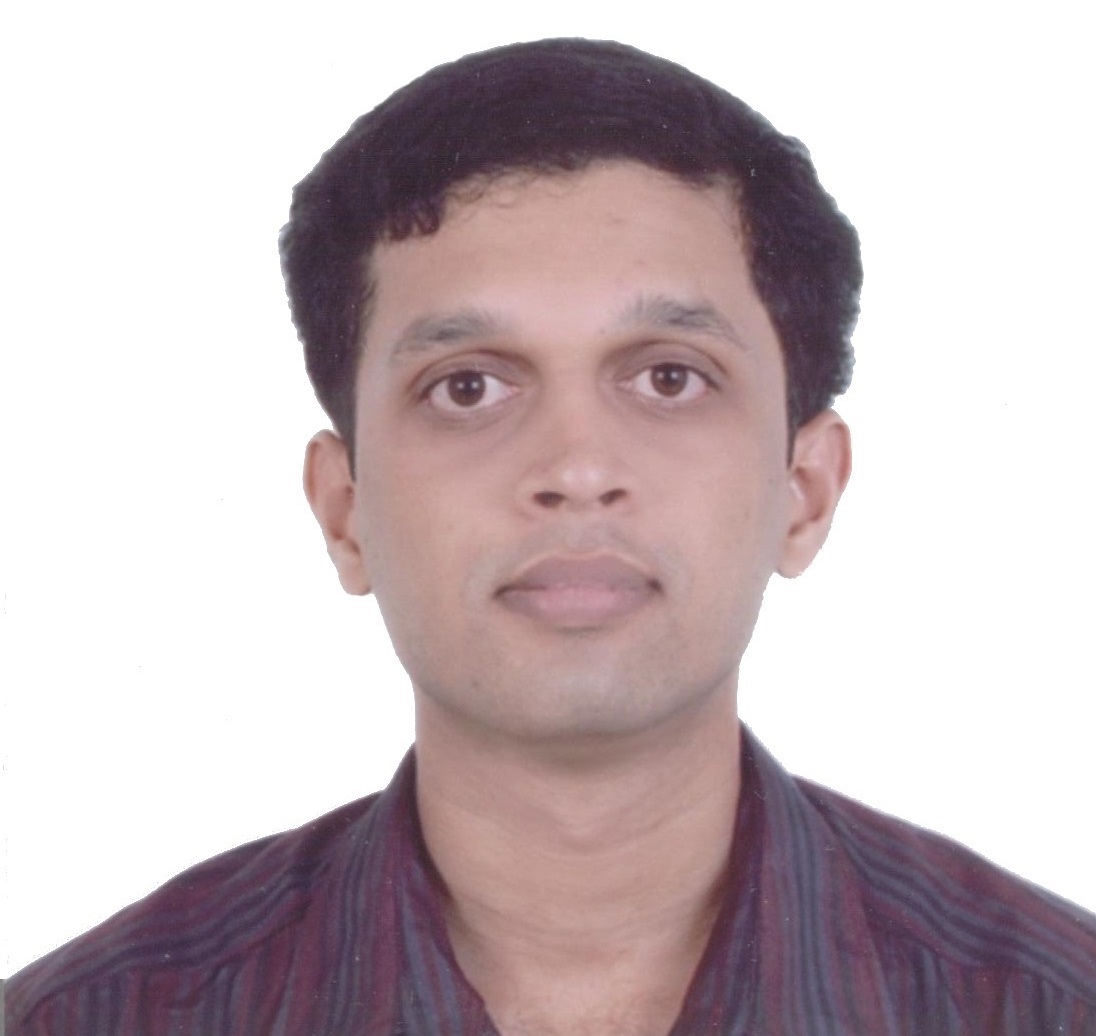}}]{Vinay C. Patil}(Student Member, IEEE)
(M\rq14) is currently pursuing his Ph.D. degree under Prof. Sandip Kundu in the Department of Electrical and Computer Engineering from the University of Massachusetts Amherst, Amherst, MA, USA. His research interests include VLSI circuit design, hardware security, machine learning and neuromorphic computing. 
\end{IEEEbiography}

\begin{IEEEbiography}[{\includegraphics[width=1in,height=1.25in,clip,keepaspectratio]{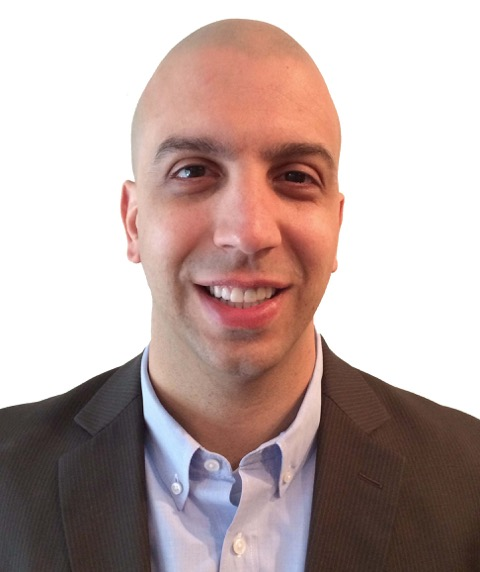}}]{Daniel E. Holcomb}(Member, IEEE)
(M\rq07) received the B.S.  and M.S.  degrees in electrical and computer engineering from the University of Massachusetts Amherst, Amherst, MA, USA, and the Ph.D.  degree in electrical engineering and computer sciences from the University of California Berkeley, Berkeley, CA, USA.  He is currently an Assistant Professor with the Department of Electrical and Computer Engineering, University of Massachusetts Amherst.  His research interests include span formal verification, very large-scale integration, embedded systems, and hardware security.
\end{IEEEbiography}

\begin{IEEEbiography}[{\includegraphics[width=1in,height=1.25in,clip,keepaspectratio]{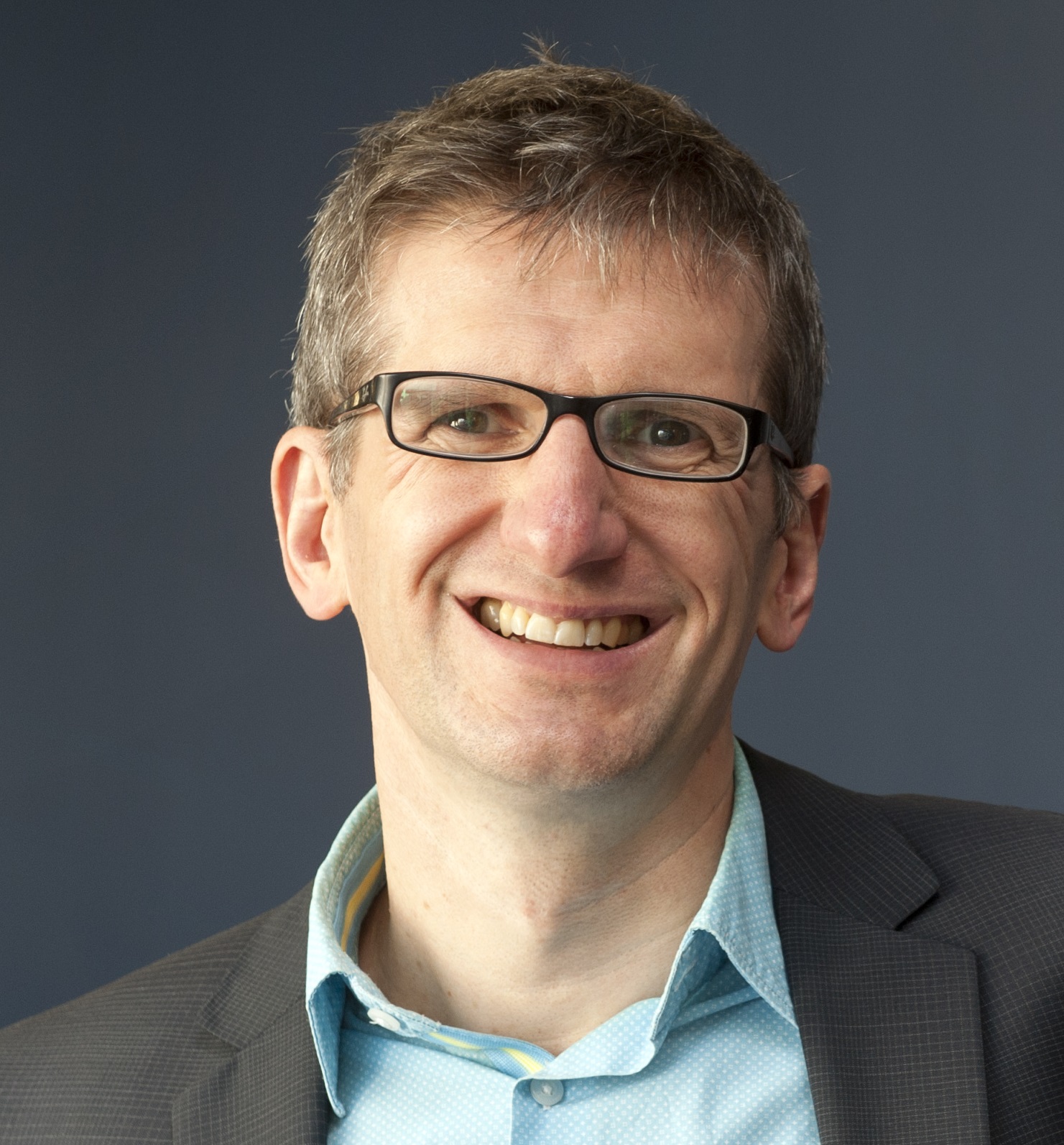}}]{Christof Paar}(Fellow, IEEE)
received the M.Sc. degree from the University of Siegen and the Ph.D. degree from the Institute for Experimental Mathematics at the University of Essen, Germany.

He holds the Chair for Embedded Security at Ruhr University Bochum, Bochum, Germany, and is an Affiliated Professor at the University of Massachusetts Amherst, Amherst, MA, USA. He cofounded, with C. Koc, the Workshop on Cryptographic Hardware and Embedded Systems (CHES) series. He has over 150 peer-reviewed publications and is coauthor of the textbook Understanding Cryptography (New York, NY, USA: Springer-Verlag, 2010). He is a cofounder of ESCRYPT -- Embedded Security, a leading consultancy firm in applied security that is now part of Bosch. His research interests include implementation techniques for cryptography, hardware security, physical security and security evaluation of real-world systems.
\end{IEEEbiography}

\begin{IEEEbiography}[{\includegraphics[width=1in,height=1.25in,clip,keepaspectratio]{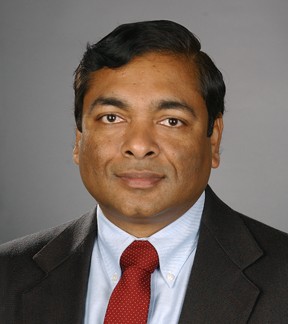}}]{Sandip Kundu}(Fellow, IEEE)
is a Professor at the University of Massachusetts at Amherst. Prior to joining academia, he spent several years in industry: first as a Research Staff Member at IBM Research Division and then at Intel Corporation as a Principal Engineer. He has published well-over 200 research papers in VLSI design and test and holds several key patents including ultra-drowsy sleep mode in processors, and has given more than a dozen tutorials at various conferences. He is a Fellow of the IEEE, Fellow of the Japan Society for Promotion of Science (JSPS), Senior International Scientist of the Chinese Academy of Sciences and a Distinguished Visitor of the IEEE Computer Society. He is currently an Associate Editor of the IEEE Transactions on Dependable and Secure Computing. Previously, he has served as an Associate Editor of the IEEE Transactions on Computers, IEEE Transactions on VLSI Systems and ACM Transactions on Design Automation of Electronic Systems. He has been Technical Program Chair/General Chair of multiple conferences including ICCD, ATS, ISVLSI, DFTS and VLSI Design Conference.
\end{IEEEbiography}

\end{document}